\documentclass[journal]{IEEEtran}
\ifCLASSINFOpdf
\else
\fi
\usepackage{nomencl}
\usepackage{amssymb}
\usepackage{float}
\usepackage{graphicx}
\usepackage{subfigure}
\usepackage{etoolbox}
\usepackage{cite}
\usepackage{amsmath,amssymb,amsfonts}
\usepackage{amsmath,textcomp}
\usepackage{algorithmic}
\usepackage{textcomp}
\usepackage{xcolor}
\usepackage{multirow}
\usepackage{makecell}
\usepackage{booktabs}
\usepackage{tipa}
\usepackage{dsfont}
\usepackage{textcomp}
\usepackage{threeparttable}
\usepackage{threeparttable}
\usepackage{upgreek}
\usepackage{mathrsfs}
\usepackage{CJKutf8}
\usepackage{xcolor}
\usepackage{adjustbox}
\usepackage{tabularx}
\makenomenclature
\begin{document}
\title{Life-Cycle Planning of Collector System for Deep-Sea Multi-Spatial Wind-PV-Tidal Farm}
\author{Wenhao~Gao,\IEEEmembership{}
        Weitai~Xu,\IEEEmembership{}
        Yunfei~Du,\IEEEmembership{}
        Yongheng~Wang,\IEEEmembership{}
        Xinwei~Shen*\IEEEmembership{}
\thanks{W. Gao, W. Xu, Y. Du and X. Shen are with Tsinghua Shenzhen International Graduate School, Tsinghua University. (Corresponding: X. Shen, sxw.tbsi@sz.tsinghua.edu.cn)}
\thanks{Y. Wang is with the Department of Electrical and Computer Engineering, The University of Hong Kong.}
\thanks{This work is supported in part by National Natural Science Foundation of China (52477102), Guangdong Basic and Applied Basic Research Foundation (2023A1515240055), Excellent Youth Basic Research Fund of Shenzhen (RCYX20231211090430053).}
}
\maketitle
\begin{abstract}
This paper develops a life-cycle optimization model for the collector system (CS) of deep-sea co-located energy farms (CEFs), where co-located energy turbines (CETs) integrate wind, photovoltaic (PV), and tidal current resources across sea-area layers. The model captures multi-layer marine-space complementarity, wake effects, and output variability, while accommodating diverse dynamic submarine cable configurations. To improve computational efficiency, the adaptive piecewise linearization (A-PWL) method based on the outputs of CETs is proposed to transform the original mixed-integer quadratic programming (MIQP) problem into a mixed-integer linear programming (MILP) form to approximate quadratic operation costs and simplify absolute power flow modeling. Case studies demonstrate that incorporating multi-energy complementarity significantly enhances the economic performance of deep-sea CEFs. When external physical risks are negligible, the fully-suspended cable configuration proves more cost-effective than the lazy-wave design. The proposed linearization method achieves high accuracy while significantly reducing solution time. Overall, this work provides a practical and scalable framework for efficient CS planning in offshore renewable energy systems.
\end{abstract}
\vspace{-1em}
\IEEEpeerreviewmaketitle
\renewcommand\nomgroup[1]{%
  \item[\bfseries
  \ifstrequal{#1}{A}{Indices and Sets}{%
  \ifstrequal{#1}{B}{Parameters}{%
  \ifstrequal{#1}{C}{Variables}{}}}%
]}
\nomenclature[A,01]{\(i,j\)}{Index of turbines.}
\nomenclature[A,02]{\(ij\)}{Index of cables.}
\nomenclature[A,03]{\(e\)}{Index of scenarios.}
\nomenclature[A,04]{\(k\)}{Index of submarine cable types.}
\nomenclature[A,05]{\(t\)}{Index of dynamic cable configurations.}
\nomenclature[A,06]{\(N/N^\mathrm{turb}/N^\mathrm{sub}\)}{Set of nodes/turbine nodes/substations.}
\nomenclature[A,07]{\(L\)}{Set of submarine cables.}
\nomenclature[A,08]{\(L^\mathrm{cross}\)}{Set of crossing submarine cables.}
\nomenclature[A,09]{\(E\)}{Set of scenarios.}
\nomenclature[A,10]{\(K\)}{Set of submarine cable types.}
\nomenclature[A,11]{\(T\)}{Set of dynamic cable configurations.}
\nomenclature[B,01]{\(c_{ij,t,k}\)}{Investment cost of submarine cable $ij$ with configuration $t$ and type $k$.}
\nomenclature[B,02]{\(\xi_{t}\)}{Margin of submarine cable with configuration $t$.}
\nomenclature[B,03]{\(d_{ij}\)}{Euclidean distance between turbines $i$ and $j$.}
\nomenclature[B,04]{\(l_{ij}^\mathrm{dyn}\)}{Length of the dynamic section of submarine cable $ij$.}
\nomenclature[B,05]{\(h_{i}\)}{Seabed depth at the location of turbine $i$.}
\nomenclature[B,06]{\(c_{t,k}^\mathrm{unit}\)}{Unit investment cost of submarine cable with configuration $t$ and type $k$.}
\nomenclature[B,07]{\(c^\mathrm{aux}\)}{Investment cost of auxiliary devices of floating cable systems.}
\nomenclature[B,08]{\(c^\mathrm{vessel}\)}{Installation cost of submarine cables.}
\nomenclature[B,09]{\(c^\mathrm{vessel,aux}\)}{Installation cost of auxiliary equipment.}
\nomenclature[B,10]{\(c^\mathrm{mobile}\)}{Mobilization cost of the installation vessels.}
\nomenclature[B,11]{\(l_{ij,t}\)}{Total length of submarine cables in configuration $t$ including both dynamic and buried static sections.}
\nomenclature[B,12]{\(c^\mathrm{cu}\)}{Cost of copper recycling.}
\nomenclature[B,13]{\(r_{ij,t,k}\)}{Cross-sectional radius of submarine cable with configuration $t$ and type $k$.}
\nomenclature[B,14]{\(c^\mathrm{ele}\)}{Grid feed-in price of offshore renewable generation.}
\nomenclature[B,15]{\(R_{ij,k}\)}{Resistance of cable $ij$ with type $k$.}
\nomenclature[B,16]{\(p_e\)}{Probability of scenario $e$.}
\nomenclature[B,17]{\(q_k\)}{Failure rate of submarine cable with type $k$.}
\nomenclature[B,18]{\(c^\mathrm{thr}\)}{Unit generation cost of thermal units.}
\nomenclature[B,19]{\(c^\mathrm{cb}\)}{Carbon tax in regional carbon market.}
\nomenclature[B,20]{\(\rho^\mathrm{cb}\)}{Carbon emission intensity of thermal units.}
\nomenclature[B,21]{\(D_i\)}{Virtual load at node $i$.}
\nomenclature[B,22]{\(\underline{n^\mathrm{feeder}}/\overline{n^\mathrm{feeder}}\)}{Lower/upper bound on the number of feeders at substation.}
\nomenclature[B,23]{\(B_{ij}\)}{Susceptance of cable $ij$.}
\nomenclature[B,24]{\(\overline{P_{ij,k}}\)}{Maximum active power capacity of submarine cable type $k$.}
\nomenclature[B,25]{\(\overline{P_{i,e}^\mathrm{pv}}\)}{Maximum active power output of PV at node $i$ under scenario $e$.}
\nomenclature[B,26]{\(\overline{P_{i}^\mathrm{sub}}\)}{Maximum active power of offshore substation (OSS) at node $i$.}
\nomenclature[C,01]{\(z_{ij}\)}{Binary variable associated with submarine cable $ij$.}
\nomenclature[C,02]{\(P_{ij,e}\)}{Active power flow in submarine cable $ij$ under scenario $e$.}
\nomenclature[C,03]{\(P_{i,e}^\mathrm{wt,cur}\)}{Curtailed wind power at node $i$ under scenario $e$.}
\nomenclature[C,04]{\(P_{i,e}^\mathrm{tct,cur}\)}{Curtailed tidal current power at node $i$ under scenario $e$.}
\nomenclature[C,05]{\(y_{ij}\)}{Binary variable associated with parent-child node relationship between nodes $i$ and $j$ in radial topology.}
\nomenclature[C,06]{\(F_{ij}\)}{Virtual power flow in submarine cable $ij$.}
\nomenclature[C,07]{\(\theta_{i,e}\)}{Voltage phase angle at node $i$ under scenario $e$.}
\nomenclature[C,08]{\(P_{i,e}^\mathrm{wt}\)}{Active power output of wind turbine at node $i$ under scenario $e$.}
\nomenclature[C,09]{\(P_{i,e}^\mathrm{tct}\)}{Active power output of tidal current turbine at node $i$ under scenario $e$.}
\nomenclature[C,10]{\(P_{i,e}^\mathrm{pv}\)}{Active power output of PV at node $i$ under scenario $e$.}
\nomenclature[C,11]{\(P_{i,e}^\mathrm{sub}\)}{Active power of OSS at node $i$ under scenario $e$.}
\nomenclature[C,12]{\(w_n\)}{Continuous auxiliary variable for A-PWL segment $n$.}
\nomenclature[C,13]{\(z_n\)}{Binary auxiliary variable for A-PWL segment $n$.}
\nomenclature[C,14]{\({P}^{+}_{ij,e}\)}{Auxiliary variable for the absolute value of active power flow in submarine cable $ij$ under scenario $e$.}

\printnomenclature[2.3cm]
\vspace{-1em}
\section{Introduction}
\IEEEPARstart{M}{ulti-layer} utilization of sea areas has recently been emphasized as an important direction for the high-quality development of the marine economy \cite{xinhua2025marineeconomy}. This policy orientation motivates a shift from conventional two-dimensional nearshore layouts to deep-sea multi-spatial infrastructure planning, where above-sea wind structures, platform-mounted PV units, and water-column tidal current turbines are coordinated within one system. As nearshore sea space becomes increasingly constrained, deep-sea renewable deployment increasingly relies on floating generation units and dynamic submarine cables. In such a setting, the collector system (CS) is no longer a conventional single-source cable-routing problem. Instead, it must coordinate multi-source power collection, floating-specific cable behavior, and deep-water life-cycle costs. Therefore, CS planning for deep-sea multi-spatial farms is substantially more challenging than conventional nearshore planning \cite{kallinger2023offshore}.

In traditional offshore wind farm (OWF) CS planning, prior research has often cast the problem as a graph-theoretic topology optimization task, with radial layouts \cite{lakshmanan2021electrical} and ring structures \cite{shen2023optimal} being the predominant configurations. Minimum spanning tree formulations \cite{dutta2012optimal} and metaheuristic methods such as particle swarm optimization \cite{hou2017overall} can identify feasible layouts efficiently, but they do not guarantee global optimality. From the mathematical optimization perspective, reliability-oriented mixed-integer programming (MIP) models \cite{ding2024smart} and loss-aware preprocessing strategies \cite{perez2019global} have progressively improved model realism. However, these studies mainly target bottom-fixed, wind-only farms, and thus do not explicitly capture floating-specific inter-array cables or the collector-level effect of multi-source complementary generation.

For deep-sea floating offshore wind farms (FOWFs), the floating nature of wind turbines (WTs) makes CS planning more involved because inter-array cables must adapt to platform motion and large water depth. Dynamic submarine cables such as lazy-wave and fully-suspended configurations \cite{kallinger2023offshore} introduce additional tradeoffs among electrical performance, mechanical integrity, and installation cost. Existing FOWF studies have considered metaheuristic inter-array planning \cite{lerch2021metaheuristic}, hybrid heuristic models under wind uncertainty \cite{song2023optimization}, and exact optimization of inter-array dynamic cable networks \cite{perezrua2024exact}. Prior work on dynamic cable configuration design has further examined lazy-wave and suspended concepts from mechanical and techno-economic viewpoints \cite{rentschler2019dynamic}. These studies substantially advance floating wind collection planning, but they remain largely wind-only or configuration-oriented rather than unified CS planning models for multi-energy deep-sea farms.

The CS planning of tidal current farms (TCFs) shares similarities with OWFs. Existing research commonly focuses on the integration of micro-siting \cite{dai2017optimal} and CS planning \cite{ren2018coordinated} under wake effects caused by tidal current turbines (TCTs). Temporal output variability and 3D seabed topography have also been incorporated in mixed-integer linear programming (MILP)-based TCF planning \cite{ren20213}. To exploit marine resource complementarity, co-located offshore wind--tidal deployment has been assessed from mechanical, civil, and economic perspectives \cite{lande2019co}, and optimized from the electrical-layout viewpoint \cite{tao2021optimal}. These studies confirm the value of offshore resource complementarity. However, they mainly focus on micro-siting, farm-level layout, or project-level feasibility assessment, while the CS-layer is still treated in an aggregated way. In particular, the joint effect of multi-source loading, cable type selection, dynamic cable configuration, and life-cycle electrical-network cost remains insufficiently modeled, and PV has rarely been incorporated into this layer.

\begin{figure}[htbp]
    \centering
    \includegraphics[width=0.98\linewidth]{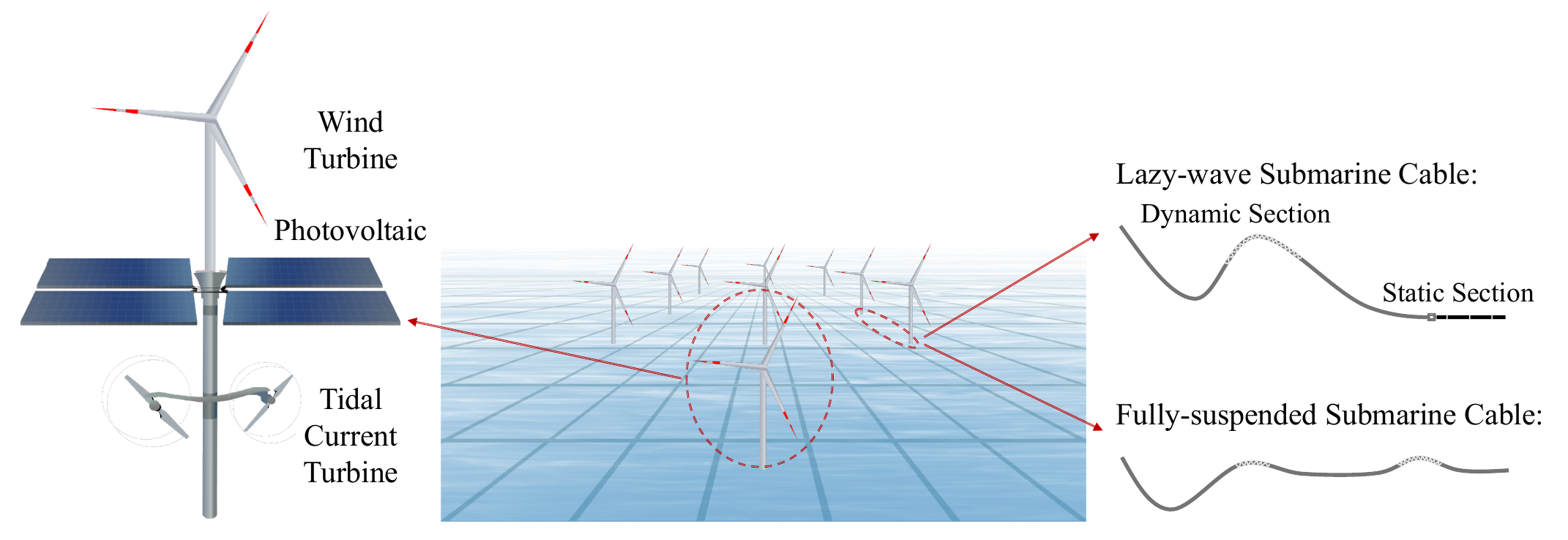}
    \caption{Schematic of the proposed CEF and CET for multi-layer marine-space utilization.}
    \label{fig_sketch}
\end{figure}

Uncertainty modeling is another necessary component because renewable outputs vary across time and space. Scenario generation has been widely studied in the power-systems literature, including wind-power scenario reviews \cite{li2020reviewscenario}, multi-renewable scenario construction with spatial and temporal correlations \cite{tang2018scenario}, and renewable-scenario platforms \cite{iversen2016resgen}. For the CS planning of deep-sea co-located energy farms (CEFs), however, the key issue is not scenario generation itself, but how to embed a compact multi-source uncertainty representation into a large-scale MIP model. Meanwhile, CS planning is a large-scale NP-hard problem \cite{cazzaro2023combined}; warm-start strategies \cite{11225152} can accelerate MIP solving, but they do not simplify the underlying model structure. Related loss preprocessing strategies \cite{perez2019global} can preserve MILP tractability, but they still rely on simplified loading representations and neglect wake-induced output heterogeneity, which limits their realism for deep-sea CEF-CS planning.

Against this background, existing studies have separately examined bottom-fixed OWF CS planning, FOWF dynamic cable optimization, TCF planning, and co-located wind--tidal deployment. What is still lacking is a unified life-cycle CS planning framework for deep-sea co-located wind--PV--tidal farms that simultaneously accounts for floating-specific dynamic cable configurations, stochastic multi-source operating states, and CS topology and cable planning. Accordingly, the novelty of this paper is not offshore co-location per se, but the explicit treatment of wind--PV--tidal co-location at the CS planning layer, where complementarity, cable loading, losses, reliability, and infrastructure cost interact most directly.

To address this gap, this paper develops a life-cycle optimization model for deep-sea CEF-CS planning, as illustrated in Fig. \ref{fig_sketch}. The main contributions are summarized as follows:

1) A life-cycle CS planning model is developed for deep-sea CEFs, covering investment, construction, maintenance, decommissioning, operation loss, reliability, and proactive curtailment costs, while accounting for alternative dynamic cable configurations associated with co-located energy turbines (CETs). This contribution extends offshore co-location research from project-level assessment to explicit CS planning.

2) A unified CS planning framework is established for co-located wind, tidal current, and PV generation, enabling CS-level comparison between single-energy and multi-energy deployment schemes and quantifying how multi-source complementarity reshapes cable loading, topology selection, and life-cycle economic performance. The importance of this contribution lies in moving the analysis from general co-location feasibility to concrete network-planning decisions.

3) An adaptive piecewise linearization (A-PWL) method is developed to reformulate quadratic operation cost terms into a tractable MILP model, thereby improving computational efficiency while preserving the key tradeoffs among cable investment, losses, and reliability. The methodological significance of this contribution is that it makes life-cycle CS planning computationally practical for larger deep-sea cases.

The remainder of this paper is organized as follows. Section II details the proposed CEF-CS planning model. Section III describes the proposed linearization method for simplifying the large-scale MIP problem. Section IV presents case studies to evaluate the effectiveness of the proposed model and linearization method. Finally, Section V concludes the paper.
\vspace{-1em}
\section{Model Formulation}
\subsection{Scenario Generation Method}

Annual wind and tidal current statistics are converted into two marginal scenario sets via Monte Carlo sampling and discretization into representative direction--speed bins. Let
\(\{(\omega_a^{\mathrm{wt}},p_a^{\mathrm{wt}})\}_{a=1}^{S}\)
and
\(\{(\omega_b^{\mathrm{tct}},p_b^{\mathrm{tct}})\}_{b=1}^{S}\)
denote the marginal wind and tidal current scenario sets, where \(a\) and \(b\) are the marginal scenario indices.
\(\omega_a^{\mathrm{wt}}\) and \(\omega_b^{\mathrm{tct}}\) are the \(a\)-th representative wind scenario and the \(b\)-th tidal current scenario, including the speed and direction of wind and current;
\(p_a^{\mathrm{wt}}\) and \(p_b^{\mathrm{tct}}\) are their marginal occurrence probabilities;
and \(S\) denotes the number of representative marginal scenarios.
Restricted by data availability, the physical time delay between wind and tidal current peaks, tidal inequality, and the actual site-specific chronological wind--tidal correlation cannot be directly identified from the annual statistics. In this study, the FH-inspired rank-matching method in Algorithm~1 is used as a compact planning-oriented representative coupling under limited synchronized measurements~\cite{fan2014copulas,feng2022distributionally}. The resulting scenario weights $p_e$ are interpreted as planning weights for evaluating cable loading, losses, and reliability, rather than as an empirical joint probability estimated from synchronized wind--tidal time-series measurements.

\begin{table}[!htbp]
\footnotesize
\centering
{
\renewcommand{\arraystretch}{1.1}
\begin{tabular}{@{}>{\raggedleft\arraybackslash}p{2em}@{\hspace{0.5em}}p{0.92\linewidth}@{}}
\Xhline{1pt}
\multicolumn{2}{@{}l}{\textbf{Algorithm 1:} \quad Wind--Tidal Joint Scenario Generation Method} \\
\hline
1: & \textbf{Input:} Wind marginal scenario probabilities \(\{p_a^{\mathrm{wt}}\}_{a=1}^{S}\), tidal marginal scenario probabilities \(\{p_b^{\mathrm{tct}}\}_{b=1}^{S}\), and the number of scenarios \(S\). \\
2: & \textbf{Sort:} Sort \(\{p_a^{\mathrm{wt}}\}_{a=1}^{S}\) and \(\{p_b^{\mathrm{tct}}\}_{b=1}^{S}\) in descending order, obtaining \((p_{(r)}^{\mathrm{wt}},a_{(r)})\) and \((p_{(r)}^{\mathrm{tct}},b_{(r)})\), for \(r=1,\ldots,S\). \\
3: & \textbf{Initialize:} \(\tilde p_r\gets0\), \(a_r\gets0\), \(b_r\gets0\), \(\forall r=1,\ldots,S\). \\
4: & \textbf{for} \(r=1,\ldots,S\) \textbf{do} \\
5: & \quad \textbf{Pair by rank:} \(a_r\gets a_{(r)}\), \(b_r\gets b_{(r)}\). \\
6: & \quad \textbf{Assign representative weight:} \(\tilde p_r\gets \min\{p_{(r)}^{\mathrm{wt}},p_{(r)}^{\mathrm{tct}}\}\). \\
7: & \textbf{end for} \\
8: & \textbf{Normalize:} \(p_r\gets \tilde p_r/\sum_{\ell=1}^{S}\tilde p_\ell\), \( \forall r=1,\ldots,S\). \\
9: & \textbf{Output:} Joint wind--tidal scenario set \(E=\{e_r=(a_r,b_r,p_r)\}_{r=1}^{S}\). \\
\hline
\end{tabular}
}
\end{table}

For clarity, the rank-matching procedure can be equivalently expressed as a sparse representative joint weight matrix:
\begin{flalign}
\widehat{\pi}_{ab}^{\mathrm{RM}}
=
\sum_{r=1}^{S}
p_r
\mathbf{1}\{a=a_r,\ b=b_r\},
\label{eq_sparse_joint_weight}
\end{flalign}
where \(\widehat{\pi}_{ab}^{\mathrm{RM}}\) is the representative joint weight assigned to wind scenario \(a\) and tidal current scenario \(b\), and \(\mathbf{1}\{\cdot\}\) is the indicator function, which equals 1 if the condition inside the braces is true and equals 0 otherwise. Thus, \(\widehat{\pi}_{ab}^{\mathrm{RM}}=p_r\) if wind scenario \(a\) and tidal current scenario \(b\) are paired in the \(r\)-th joint scenario; otherwise, \(\widehat{\pi}_{ab}^{\mathrm{RM}}=0\). Since only the \(S\) rank-matched pairs have nonzero weights, \(\widehat{\pi}_{ab}^{\mathrm{RM}}\) is sparse, yielding a compact representative planning weight matrix constructed from the available marginal scenario probabilities. This treatment allows the CS planning model to be solved with a limited number of representative joint scenarios when only marginal annual wind and tidal current statistics are available.

If synchronized wind--tidal measurements become available, the same CS planning framework can directly accommodate a time-lag-aware empirical joint distribution by changing the scenario-construction input \((E,p_e)\). Let \(N^{\mathrm{obs}}\) denote the number of synchronized observations, and let \(v^{\mathrm{wt}}(t)\) and \(v^{\mathrm{tct}}(t)\) denote the wind speed and tidal current speed at observation index \(t\), respectively. For a candidate integer lag \(\tau \in \mathcal{T}=\{0,1,\ldots,\tau^{\max}\}\), the lagged wind--tidal correlation can be compactly written as
\begin{flalign}
\rho(\tau)
=
\operatorname{corr}
\left(
v^{\mathrm{wt}}(t),
v^{\mathrm{tct}}(t+\tau)
\right),
\quad
\forall \tau \in \mathcal{T},
\label{eq_lagged_corr_compact}
\end{flalign}
where \(\rho(\tau)\) is calculated over all valid overlapping samples \(t=1,\ldots,N^{\mathrm{obs}}-\tau\), and \(\tau^{\max}<N^{\mathrm{obs}}\). The dominant statistical time lag can be estimated by
\begin{flalign}
\tau^\star
\in
\arg\max_{\tau \in \mathcal{T}} \,
\left|
\rho(\tau)
\right|,
\label{eq_dominant_lag_compact}
\end{flalign}
where \(\tau^\star\) is one lag that gives the strongest absolute wind--tidal correlation. If several lags give the same maximum value, the smallest lag can be selected. Negative lags can be tested analogously by shifting the wind speed series after the tidal current series.
After estimating \(\tau^\star\), the lag-corrected empirical joint probability can be obtained by frequency counting. Let \(\omega^{\mathrm{wt}}(t)\) and \(\omega^{\mathrm{tct}}(t)\) denote the representative wind and tidal scenario assigned to the speed--direction observation at index \(t\), respectively. Then the empirical joint probability of wind scenario \(a\) and lag-corrected tidal current scenario \(b\) is

\begin{align}
\pi_{ab}^{\mathrm{emp}}(\tau^\star)
=
\frac{1}{N^{\mathrm{obs}}-\tau^\star}
\sum_{t=1}^{N^{\mathrm{obs}}-\tau^\star}
\mathbf{1}
\big\{
\omega^{\mathrm{wt}}(t)=\omega_a^{\mathrm{wt}},
\notag \\
\omega^{\mathrm{tct}}(t+\tau^\star)=\omega_b^{\mathrm{tct}}
\big\}.
\label{eq_empirical_joint_probability_compact}
\end{align}

The numerator in \eqref{eq_empirical_joint_probability_compact} counts the number of valid observations in which the wind state is assigned to scenario \(a\) and the lag-corrected tidal current state is assigned to scenario \(b\), while the denominator \(N^{\mathrm{obs}}-\tau^\star\) is the number of valid overlapping observations after applying the lag correction. By construction,
\(\pi_{ab}^{\mathrm{emp}}(\tau^\star)\ge0\) and
\(\sum_{a=1}^{S}\sum_{b=1}^{S}\pi_{ab}^{\mathrm{emp}}(\tau^\star)=1\).
Thus, \(\pi_{ab}^{\mathrm{emp}}(\tau^\star)\) is a valid empirical joint probability distribution, which can capture the chronological wind--tidal dependence when synchronized measurements are available. Tidal inequality can be further considered by constructing such empirical joint probability matrices separately for flood/ebb periods or spring/neap periods. \eqref{eq_lagged_corr_compact}--\eqref{eq_empirical_joint_probability_compact} provide a physically interpretable way to incorporate wind--tidal time delay when synchronized measurements become available ~\cite{ghaffari2010acoustic,moreira2009tidal}.
\vspace{-1em}
\subsection{Wake Effect Model}
In the deep-sea OWF, downstream turbines are typically affected by the wake effects of upstream turbines, resulting in reduced wind speed and consequently lower power output. To accurately estimate the wake loss of each turbine, this study adopts the 2D Jensen wake model \cite{ge2019two}. The wake loss $\delta_{ij}$ imposed by the turbine located at node $i$ on the turbine at node $j$ can be calculated using the following equation:
\begin{flalign}
\delta_{ij}=\frac{2\left(1-\sqrt{1-C_\mathrm{T}}\right)}{\left(\varepsilon x_{ij} / r_\mathrm{a}+1\right)^2} \exp \left[-\frac{2}{\left(\varepsilon x_{ij} / r_\mathrm{a}+1\right)^2}\left(\frac{r}{r_\mathrm{a}}\right)^2\right],
\label{eq_Wake1}
\end{flalign}
where $C_\mathrm{T}$ is the thrust coefficient, $\varepsilon$ is the wake expansion rate, $x_{ij}$ denotes the distance between turbines $i$ and $j$ along the wind direction, $r_\mathrm{a}$ is the rotor radius of WT, and $r$ is the radial distance from the rotor axis.

The root-sum-square method is applied to model the cumulative wake effect based on the momentum theory \cite{turner2014new}. Accordingly, the effective wind speed $v_j$ entering turbine $j$ is given by:
\vspace{-1.0em}
\begin{flalign}
v_j=v_\mathrm{0}\left[1-\sqrt{\sum_{i \in N^\mathrm{turb}} \delta_{ij}^2}\right],
\label{eq_Wake2}
\end{flalign}
where $N^\mathrm{turb}$ is the set of all turbine node locations, and $v_\mathrm{0}$ denotes the freestream wind speed.

Based on the effective wind speed derived above, the actual active power output $P_j$ of the $j$-th WT can be further calculated as:
\vspace{-0.2em}
\begin{flalign}
P_j= \begin{cases}P^\mathrm{rated} \cdot\frac{(v_j)^3-(v^\mathrm{in})^3}{(v^\mathrm{rated})^3-(v^\mathrm{in})^3} & v^\mathrm{in} \leq v_j<v^\mathrm{rated} \\ P^\mathrm{rated} & v^\mathrm{rated} \leq v_j<v^\mathrm{out} \\ 0 & \text { Otherwise }\end{cases},
\label{eq_Wake3}
\end{flalign}
where $v^\mathrm{in}$, $v^\mathrm{rated}$, and $v^\mathrm{out}$ represent the cut-in, rated, and cut-out wind speed of WT, respectively, and $P^\mathrm{rated}$ denotes the rated power output of the turbine.

The wake effect of tidal current is modeled using the same framework, with key parameters adjusted to capture the actual TCT output under different scenarios \cite{yang2024review}.
\vspace{-1.5em}
\subsection{Seabed Topography Model}
When investing and constructing a deep-sea CEF, the complex 3D seabed topography is required to be taken into account, as variations in seabed elevation will increase the construction cost and installation complexity. The seabed topography in this study is modeled as follows:
\begin{flalign*}
&Z(X,Y)=\overline{h}+\bigg\{ z_\mathrm{\alpha}(1-X)^2 \exp \left[-\left(X^2+Y^2\right)\right] \notag \\
&-z_\mathrm{\gamma}\left(\frac{X}{z_\mathrm{\beta}}-X^3-Y^5\right) \exp \left[-\left(X^2+Y^2\right)\right]-\underline{Z_\mathrm{p}}\bigg\}
\label{eq_seabed}
\end{flalign*}
\vspace{-1.5em}
\begin{flalign}
&&\left(\overline{h} -\underline{h}\right) /\left(\overline{Z_\mathrm{p}}-\underline{Z_\mathrm{p}}\right),
\end{flalign}
where $(X, Y, Z)$ denotes the 3D coordinates of the CEF. Parameters $z_\mathrm{\alpha}$, $z_\mathrm{\beta}$, and $z_\mathrm{\gamma}$ are tunable coefficients for generating the seabed surface. The elevation range is constrained by $\underline{h}$ and $\overline{h}$ to keep terrain variation within acceptable engineering limits. $\underline{Z_\mathrm{p}}$ and $\overline{Z_\mathrm{p}}$ represent the unnormalized elevation bounds prior to scaling.

\vspace{-1em}
\subsection{Objective Function}
The objective function of the CS consists of four components: cost of civil engineering, cost of operation, cost of reliability and cost of proactive curtailment. For consistency and ease of comparison, all components are converted into annualized costs. The overall objective function is given by:
\vspace{-0.5em}
\begin{flalign}
\min _{z,p} \;C^\mathrm{Civil}+C^\mathrm{Ope}+C^\mathrm{EENS}+C^\mathrm{Cur}.
\label{eq_Obj}
\end{flalign}

\noindent \emph{1) Cost of Civil Engineering:}
\begin{flalign}
C^\mathrm{Civil} = C^\mathrm{inv}+C^\mathrm{con}+C^\mathrm{maint}+C^\mathrm{dec},
\label{eq_Civ_1}
\end{flalign}
\vspace{-1.5em}
\begin{flalign}
C^\mathrm{inv}=CRF \sum_{k \in K} \sum_{ij \in L} c_{ij,t,k} z_{ij,t,k}, \quad \forall t \in T,
\label{eq_Civ_2}
\end{flalign}
\vspace{-1.5em}
\begin{flalign*}
c_{ij,t,k}=\left[\xi_t d_{ij}+2\left(l^\mathrm{dyn}_{ij}-h_i-h_j\right)\right]c_{t,k}^\mathrm{unit}+2c^\mathrm{aux},
\label{eq_Civ_3}
\end{flalign*}
\vspace{-2em}
\begin{flalign}
&&t=1,
\end{flalign}
\vspace{-1.5em}
\begin{flalign}
c_{ij,t,k}=\xi_t d_{ij} c^\mathrm{unit}_{t,k}, \quad t=2,
\label{eq_Civ_4}
\end{flalign}
\vspace{-1.5em}
\begin{flalign*}
C^\mathrm{con}=CRF\sum_{ij \in L} c^\mathrm{vessel} r_\mathrm{c} l_{ij,t} z_{ij,t}+c^\mathrm{vessel,aux}+c^\mathrm{mobile},
\label{eq_Civ_6}
\end{flalign*}
\vspace{-2em}
\begin{flalign}
&& \forall t \in T,
\end{flalign}
\vspace{-1.5em}
\begin{flalign}
C^\mathrm{maint}=k_\mathrm{m} C^\mathrm{inv},
\label{eq_Civ_7}
\end{flalign}
\vspace{-1.5em}
\begin{flalign*}
C^\mathrm{dec}=k_\mathrm{d} C^\mathrm{con}-CRF \sum_{k \in K} \sum_{ij \in L} c^\mathrm{cu} \rho^\mathrm{cu} \pi (r_{ij,t,k})^2 l_{ij,t} z_{ij,t,k},
\label{eq_Civ_8}
\end{flalign*}
\vspace{-2em}
\begin{flalign}
&& \forall t \in T,
\end{flalign}
\vspace{-1.5em}
\begin{flalign}
CRF=\frac{\rho(1+\rho)^\mathrm{year}}{(1+\rho)^\mathrm{year}-1}.
\label{eq_Civ_9}
\end{flalign}

The objective function \eqref{eq_Civ_1} minimizes the cost of civil engineering, which includes investment cost, construction cost, maintenance cost, and decommissioning cost. Equation \eqref{eq_Civ_2} represents the investment cost of the CS, where $c_{ij,t,k}$ denotes the investment cost of the submarine cable between nodes $i$ and $j$ under configuration $t$ and cable type $k$, and $z_{ij,t,k}$ is the corresponding binary decision variable indicating whether this cable option is selected. The capital recovery factor $CRF$ is applied to annualize the one-time investment cost over the system’s lifetime, where $\rho$ denotes the inflation rate, and $\mathrm{year}$ represents the expected operational lifespan of the CS. Expressions \eqref{eq_Civ_3}–\eqref{eq_Civ_4} correspond to the investment costs of submarine cables under lazy-wave and fully-suspended configurations\cite{lerch2021metaheuristic}. $\xi_t$ denotes the design margin under configuration $t$, $d_{ij}$ is the Euclidean distance between turbines, $l^\mathrm{dyn}_{ij}$ is the length of the dynamic section of the submarine cable, and $h_i$ is the seabed depth at the location of turbine $i$. $c^\mathrm{unit}_{t,k}$ denotes the unit investment cost of the submarine cable under the configuration and type, while $c^\mathrm{aux}$ represents the investment cost of auxiliary devices such as floating supports, bend stiffeners, and connectors for floating cable systems. Equation \eqref{eq_Civ_6} represents the construction cost of the CS, where $c^\mathrm{vessel}$, $c^\mathrm{vessel,aux}$, and $c^\mathrm{mobile}$ denote the installation cost of submarine cables, the installation cost of auxiliary devices, and the mobilization cost of the installation vessel, respectively. $r_\mathrm{c}$ is the installation cost coefficient for submarine cables, and $l_{ij,t}$ denotes the total length of the submarine cable under the configuration, including both the floating and trenched static segments. Equations \eqref{eq_Civ_7} and \eqref{eq_Civ_8} define the maintenance cost and decommissioning cost of the CS, where $k_\mathrm{m}$ and $k_\mathrm{d}$ are the cost conversion factors for maintenance and decommissioning. $c^\mathrm{cu}$ denotes the unit recycling cost of copper, and $\rho^\mathrm{cu}$ is the density of copper material. $r_{ij,t,k}$ represents the cross-sectional radius of the submarine cable under the configuration and type.

\noindent \emph{2) Cost of Operation:}
\begin{flalign}
C^\mathrm{Ope}=c^\mathrm{ele} AUH \sum_{e \in E} \sum_{k \in K} \sum_{ij \in L} P_{ij,e}^2 R_{ij,k}\, p_e.
\label{eq_Ope}
\end{flalign}

The objective function \eqref{eq_Ope} represents the operation cost of the CS resulting from power losses. In this equation, $c^\mathrm{ele}$ denotes the feed-in tariff for CEF generation. $AUH$ (annual utilization hours) represents the average annual utilization hours of the CEF system. $P_{ij,e}$ is the active power flow through submarine cable $ij$ under scenario $e$, and $R_{ij,k}$ denotes the electrical resistance of the cable under type $k$. $p_e$ is the probability of occurrence of scenario $e$ \cite{shen2023optimal}.

\noindent \emph{3) Cost of Reliability:}
\begin{flalign}
C^\mathrm{EENS} =c^\mathrm{ele} MTTR \sum_{e \in E} \sum_{k \in K} \sum_{ij \in L} l_{ij} q_k z_{ij,k} \left|P_{ij,e}\right| \,p_e.
\label{eq_EENS}
\end{flalign}

The expression \eqref{eq_EENS} represents the expected energy not served (EENS), defining the reliability cost of the CEF-CS. $MTTR$ denotes the mean time to repair, and $q_k$ is the failure rate of the cable when type $k$ is selected \cite{shen2023optimal}.

\noindent \emph{4) Cost of Proactive Curtailment:}
\begin{flalign}
C^\mathrm{Cur} = C^\mathrm{gen}+C^\mathrm{cb},
\label{eq_Cur1}
\end{flalign}
\vspace{-1em}
\begin{flalign}
C^\mathrm{gen} = c^\mathrm{thr} AUH \sum_{e \in E} \sum_{i \in N^\mathrm{turb}} (P_{i,e}^\mathrm{wt,cur}+P_{i,e}^\mathrm{tct,cur}) \,p_e,
\label{eq_Cur2}
\end{flalign}
\vspace{-1em}
\begin{flalign}
C^\mathrm{cb} = c^\mathrm{cb} \rho^\mathrm{cb} AUH \sum_{e \in E} \sum_{i \in N^\mathrm{turb}} (P_{i,e}^\mathrm{wt,cur}+P_{i,e}^\mathrm{tct,cur}) \,p_e.
\label{eq_Cur3}
\end{flalign}

Equations \eqref{eq_Cur1}–\eqref{eq_Cur3} define the cost of proactive curtailment for the CEF system. This cost consists of two components: (i) the generation cost $C^\mathrm{gen}$ incurred by thermal units to compensate for curtailed wind and tidal current energy, and (ii) the carbon emission cost $C^\mathrm{cb}$ associated with thermal generation. $c^\mathrm{thr}$ denotes the unit generation cost of thermal units, $P_{i,e}^\mathrm{wt,cur}$ and $P_{i,e}^\mathrm{tct,cur}$ represent the curtailed wind and tidal current power at turbine $i$ under scenario $e$, $c^\mathrm{cb}$ is the carbon tax in the regional carbon market, and $\rho^\mathrm{cb}$ denotes the carbon emission intensity of thermal units. In conventional CS planning models, curtailment is typically treated as a penalty term to ensure maximum utilization of the resources of offshore renewable energy. Taking OWF as an example, the curtailment cost is often defined as $C^\mathrm{Cur} = M \sum_{e \in E} \sum_{i \in N^\mathrm{turb}} P_{i,e}^\mathrm{wt,cur} \,p_e$, where a sufficiently large constant $M$ is used as the penalty coefficient to enforce zero passive curtailment, ensuring that all generated wind power is delivered through the submarine cables. In contrast, the proactive curtailment cost proposed in this study allows controlled curtailment of wind and tidal current power, which enhances the utilization of high-voltage transmission lines and transformer loading in deep-sea CEF.
\vspace{-1.0em}
\subsection{Constraints}
\noindent (1) \textit{Radial Topology Constraints:}
\begin{flalign}
\sum_{ij \in L} z_{ij}=|N^\mathrm{turb}|,
\label{eq_topology1}
\end{flalign}
\vspace{-1.5em}
\begin{flalign}
y_{ij}+y_{ji}=z_{ij}, \quad \forall ij\in L,
\label{eq_topology2}
\end{flalign}
\vspace{-1.5em}
\begin{flalign}
y_{ij}=0, \quad \forall i\in N^\mathrm{sub}, ij\in L,
\label{eq_topology3}
\end{flalign}
\vspace{-1.5em}
\begin{flalign}
\sum_{ij\in L} y_{ij}=1, \quad \forall i\in N^\mathrm{turb},
\label{eq_topology4}
\end{flalign}
\vspace{-1.5em}
\begin{flalign}
\sum_{ij\in L} F_{ij}+D_i=\sum_{ki\in L} F_{ki}, \quad \forall i\in N,
\label{eq_topology5}
\end{flalign}
\vspace{-1.5em}
\begin{flalign}
\left|F_{ij}\right| \leq z_{ij} M, \quad \forall ij\in L.
\label{eq_topology6}
\end{flalign}

Cons. \eqref{eq_topology1}–\eqref{eq_topology6} are radial topology constraints of the CS \cite{wang2023joint}. Cons. \eqref{eq_topology1} ensures that the total number of submarine cables equals the number of turbines excluding offshore substations. Cons. \eqref{eq_topology2}–\eqref{eq_topology4} impose the spanning tree structure: specifically, cons. \eqref{eq_topology2} enforces the parent–child relationship such that if node $j(i)$ is the parent of node $i(j)$, the corresponding auxiliary binary variable $y_{ij}$ ($y_{ji}$) is set to 1. Cons. \eqref{eq_topology3} and \eqref{eq_topology4} require that offshore substations have no parent node, while each turbine node must have exactly one parent.
Cons. \eqref{eq_topology5}–\eqref{eq_topology6} denote the single-commodity flow constraints, where $F_{ij}$ denotes the fictitious flow along the cable, and $D_i$ is the fictitious demand of turbine $i$. Cons. \eqref{eq_topology5} ensures fictitious flow balance, while \eqref{eq_topology6} forces the fictitious flow to zero if the cable is not selected.

\noindent (2) \textit{Submarine Cable Selection Constraint:}
\begin{flalign}
z_{i j}=\sum_{k \in K} z_{ij,k}, \quad \forall i j \in L.
\label{eq_type}
\end{flalign}

In cons. \eqref{eq_type}, each selected submarine cable must be assigned exactly one type.

\noindent (3) \textit{Cable Crossing Avoidance Constraint:}
\begin{flalign}
z_{ij}+z_{gh} \leq 1, \quad \forall(ij,gh) \in L^\mathrm{cross}.
\label{eq_crossing}
\end{flalign}

Cons. \eqref{eq_crossing} imposes the non-crossing constraint for submarine cables. $L^\mathrm{cross}$ denotes the set of cable pairs with potential crossings. For any two cables $ij$ and $gh$ belonging to $L^\mathrm{cross}$, at most one can be selected for construction \cite{shen2023optimal}.

\noindent (4) \textit{Offshore Substation-related Constraint:}
\begin{flalign}
\underline{n^\mathrm{feeder}} \leq \sum_{ij \in L} z_{ij} \leq \overline{n^\mathrm{feeder}}, \quad \forall i \in N^\mathrm{sub},
\label{eq_feeders}
\end{flalign}
\vspace{-1em}
\begin{flalign}
0 \leq P_{i,e}^\mathrm{sub} \leq \overline{P_{i}^\mathrm{sub}}, \quad \forall i \in N^\mathrm{sub}, e \in E.
\label{eq_maxpower}
\end{flalign}

Cons. \eqref{eq_feeders} limits the number of outgoing feeders from offshore substations (OSSs), and cons. \eqref{eq_maxpower} imposes an upper limit on the maximum receivable power of OSSs. $\underline{n^\mathrm{feeder}}$ and $\overline{n^\mathrm{feeder}}$ represent the minimum and maximum allowable number of feeder connections, $P_{i,e}^\mathrm{sub}$ and $\overline{P_{i}^\mathrm{sub}}$ denote the active power received by the OSS and its upper power limit.

\noindent (5) \textit{Power Flow Constraints:}
\begin{flalign*}
\left|B_{ij}\left(\theta_{j,e}-\theta_{i,e}\right)-P_{ij,e}\right| \leq\left(1-z_{ij}\right) M,
\label{PowerFlow1}
\end{flalign*}
\vspace{-2em}
\begin{flalign}
&&\forall i j \in L, e \in E,
\end{flalign}
\vspace{-2em}
\begin{flalign}
\theta_{i,e}=0, \quad \forall i \in N^\mathrm{sub}, e \in E,
\label{PowerFlow2}
\end{flalign}
\vspace{-2em}
\begin{flalign*}
\sum_{ij \in L} P_{ij,e}-\sum_{ji \in L} P_{ji,e}=P_{j,e}^\mathrm{wt}+P_{j,e}^\mathrm{tct}+P_{j,e}^\mathrm{pv}-P_{j,e}^\mathrm{wt,cur}
\label{PowerFlow3}
\end{flalign*}
\vspace{-1.5em}
\begin{flalign}
&&-P_{j,e}^\mathrm{tct,cur}-P_{j,e}^\mathrm{sub}, \quad \forall j \in N, e \in E.
\end{flalign}

Cons. \eqref{PowerFlow1}-\eqref{PowerFlow3} are the constraints of power flow, where \eqref{PowerFlow1} and \eqref{PowerFlow2} represent a planning-stage DC power flow approximation, capturing the relationship between the active power flow on cable $ij$ under different scenarios and the voltage phase angles at the corresponding turbine nodes \cite{shen2023optimal,perez2022reliability}. $B_{ij}$ denotes the susceptance of submarine cable $ij$, and $\theta_{i,e}$ is the voltage phase angle at CET $i$ in scenario $e$. The OSS is defined as the reference node, and its phase angle is fixed at zero. The active power balance is described by cons. \eqref{PowerFlow3}. $P_{j,e}^\mathrm{wt}$, $P_{j,e}^\mathrm{tct}$, and $P_{j,e}^\mathrm{pv}$ represent the active power outputs from WT, TCT, and PV at node $j$ under each scenario, respectively. To write the active power balance compactly over the whole node set $N$, the quantities associated with physically absent devices are defined as zero. For every OSS node $j\in N^\mathrm{sub}$, there is no local WT, TCT, or PV generation, hence $P_{j,e}^\mathrm{wt}=P_{j,e}^\mathrm{tct}= {P}_{j,e}^\mathrm{pv}=P_{j,e}^\mathrm{wt,cur}=P_{j,e}^\mathrm{tct,cur}=0$, and set $P_{j,e}^\mathrm{sub}=0$ for every turbine node $j\in N^\mathrm{turb}$. Under these definitions,  \eqref{PowerFlow3} represents the net renewable injection at CET nodes and reduces to the negative of the received active power at OSS nodes. 

\noindent (6) \textit{Submarine Cable Capacity Constraint:}
\begin{flalign}
\left|P_{ij,e}\right| \leq \sum_{k \in K} z_{ij,k} \overline{P_{ij,k}}, \quad \forall i j \in L, e \in E.
\label{eq_Capacity}
\end{flalign}

Cons. \eqref{eq_Capacity} ensures that the active power flow through each submarine cable does not exceed its maximum allowable capacity $\overline{P_{ij,k}}$ under the selected cable type $k$.

\noindent (7) \textit{CET Operational Constraints:}
\begin{flalign}
0 \leq P_{i,e}^\mathrm{wt,cur} \leq P_{i,e}^\mathrm{wt}, \quad \forall i \in N^\mathrm{turb}, e \in E,
&\label{eq_Ope1}
\end{flalign}
\vspace{-1em}
\begin{flalign}
0 \leq P_{i,e}^\mathrm{tct,cur} \leq P_{i,e}^\mathrm{tct}, \quad \forall i \in N^\mathrm{turb}, e \in E,
&\label{eq_Ope2}
\end{flalign}
\vspace{-1em}
\begin{flalign}
0 \leq P_{i,e}^\mathrm{pv} \leq \overline{P_{i,e}^\mathrm{pv}}, \quad \forall i \in N^\mathrm{turb}, e \in E.
&\label{eq_Ope3}
\end{flalign}

The CET operation is constrained by \eqref{eq_Ope1}–\eqref{eq_Ope3}, ensuring that WT/TCT curtailment does not exceed their generation and PV output does not exceed its available power. 
Since PV fluctuation is mainly intra-day and the installed PV capacity at each CET is relatively small compared with the co-located wind and tidal current capacities, the available PV power in the base planning model is set to its rated value as a conservative cable-loading assumption for life-cycle CS sizing.
\vspace{-1.0em}
\subsection{Baseline Costs for CEF}
As a metric for evaluating and comparing the life-cycle costs of the proposed CEF system, the levelized cost of energy (LCOE) is typically defined as \cite{vazquez2015device}:
\begin{flalign}
LCOE=\frac{\sum_\tau\left[\left(CAPEX_\tau+OPEX_\tau\right)(1+\rho)^{-\tau}\right]}{\sum_\tau\left[E_\tau(1+\rho)^{-\tau}\right]},
\label{eq_LCOE}
\end{flalign}
where $CAPEX_\tau$ denotes the capital expenditure in year $\tau$, representing the annualized investment cost of the CEF system. $OPEX_\tau$ refers to the operational cost in year $\tau$, and $E_\tau$ is the total amount of energy generated during the year.
\section{Adaptive Piecewise Linearization for Simplifying Large-Scale MIP}
\subsection{Adaptive Linearization Method of the Proposed Model}
The proposed mathematical model can be formulated as a large-scale mixed-integer quadratic programming (MIQP) problem. To simplify the problem, the A-PWL method is developed building on the work in \cite{keha2004models}. The nonlinear term $P_{ij, e}^2$ is linearized over its domain $\left[p_1, p_{n+1}\right]$ with $n+1$ breakpoints selected according to the rated output of each CET (WT\&TCT\&PV). Accordingly, $n+1$ continuous variables $\left[w_1, w_2, \cdots, w_{n+1}\right]$ and $n$ binary variables $\left[z_1, z_2, \cdots, z_n\right]$ are introduced, subject to:
\vspace{-0.5em}
\begin{flalign}
\sum_{i=1}^{n+1} w_i=1,
&\label{eq_Linear1}
\end{flalign}
\vspace{-1em}
\begin{flalign}
\sum_{i=1}^n z_i=1,
&\label{eq_Linear2}
\end{flalign}
\vspace{-1em}
\begin{flalign}
w_1 \geq 0, w_2 \geq 0, \cdots, w_{n+1} \geq 0,
&\label{eq_Linear3}
\end{flalign}
\vspace{-1em}
\begin{flalign}
w_1 \leq z_1, w_2 \leq z_1+z_2, \cdots, w_{n+1} \leq z_n.
&\label{eq_Linear4}
\end{flalign}

The original nonlinear function can then be replaced by the following linear expression:
\begin{flalign}
\left|P_{ij,e}\right|=\sum_{i=1}^{n+1} w_i p_i,
&\label{eq_Linear5}
\end{flalign}
\vspace{-1em}
\begin{flalign}
P_{ij,e}^2=\sum_{i=1}^{n+1} w_i f\left(p_i\right).
&\label{eq_Linear6}
\end{flalign}

For example, if a cable collects up to 6 CETs with a rated output of 15 MW each, $P_{ij,e}^2$ is linearized as the piecewise-linear function of $\lvert P_{ij,e}\rvert$ on $[0,90]$ MW, with breakpoints placed at 15 MW intervals. The detailed proof of the A-PWL method can be found in Table~\ref{Linear}.

\begin{table}[htbp]
\caption{Mathematical Proof of A-PWL Method}
\label{Linear}
\scriptsize
\centering
\begin{tabular}{ll} 
\hline \hline
\textbf{P-I:} 
& \textbf{Feasibility of Single-Interval Linear Interpolation:}\\
& \textbf{Assume} binary variables $\sum_{i=1}^{n} z_i$, which ensures exactly one\\
& segment index $k$ satisfies $z_k = 1$ and all other $z_j=0$.\\
& \textbf{Impose} $w_1 \le z_1, w_2 \le z_1 + z_2,\dots,w_{n+1} \le z_n$,\\
& along with the constraints $\sum_{i=1}^{n+1} w_i = 1,w_i \ge 0$, because\\
& $z_k=1$ for a unique $k$ and $z_j=0$ otherwise, only $w_k$ and\\
& $w_{k+1}$ can be nonzero.\\
& Yielding $P_{ij,e}=w_k p_k +w_{k+1} p_{k+1}$ and $w_k + w_{k+1} = 1$.\\
& \textbf{If} let $\alpha = w_k$, it follows that:\\
& $\qquad \qquad P_{i j, e}=\alpha p_k+(1-\alpha) p_{k+1}, \; \alpha \in[0,1].$ \\
& \textbf{Thus}, $P_{ij,e}$ remains within the segment $[p_k, \, p_{k+1}]$ for any\\
& desired $P_{ij,e}\in [p_{1},\,p_{n+1}]$, there is a unique segment can be \\
& expressed as a convex combination of two adjacent breakpoints.\\
& \textbf{Therefore}, the constraints \eqref{eq_Linear1}-\eqref{eq_Linear6} guarantee the feasibility of\\
& modeling $P_{ij,e}$ with single-segment linear interpolation.\\
\hline
\textbf{P-II:} 
& \textbf{Convexity and Error Bound of the Operation Cost Model:}\\
& The true quadratic value is $\bigl(\alpha \, p_k + (1-\alpha)\,p_{k+1}\bigr)^2$, while\\
& the A-PWL approximation is $\alpha\, p_k^2 + (1-\alpha)\, p_{k+1}^2$.\\
& \textbf{Since} $f(p) = p^2$ is convex, \textbf{Jensen's inequality} gives:\\
& $ \qquad \alpha\,p_k^2 + (1-\alpha)\,p_{k+1}^2 \ge\bigl(\alpha\,p_k + (1-\alpha)\,p_{k+1}\bigr)^2$.\\
& \textbf{Thus}, the linear approximation overestimates the true cost,\\
& with error:\\
& $\qquad \qquad \qquad e = \alpha(1-\alpha) \,\bigl(p_{k+1} - p_k\bigr)^2$. \\
& \textbf{Since} $\alpha(1-\alpha)\leq 1/4$ for $\alpha \in [0,1]$, the maximum error is:\\
& \qquad\qquad\qquad\quad \quad  $e^{\max} = \frac{(p_{k+1}-p_k)^2}{4}$.\\
& \textbf{Hence}, the linear interpolation provides an upper bound on the\\
& true quadratic cost, with the bound determined solely by the\\
& segment width.\\
\hline
\textbf{P-III:} 
& \textbf{Impact of Segment Refinement on Approximation Accuracy:}\\
& To span the interval $[p_1,p_{n+1}]$, divide it into $n$ segments of\\
& length $\Delta = (p_{n+1}-p_1)/{n}$ according to rated power output\\
& of CETs. Each segment introduces a maximum error of $\Delta^2 /4$.\\
& As $n$ grows, $\Delta$ decreases, and the total error diminishes at a\\
& rate of $O(1/n^2)$. \\
& \textbf{Therefore} by increasing the number of segments, the A-PWL\\
& model can approximate the original quadratic cost with\\
& arbitrarily high precision.\\

\hline \hline 
\end{tabular}
\end{table}

\begin{figure}[htbp]
    \centering
    \includegraphics[width=0.95\linewidth]{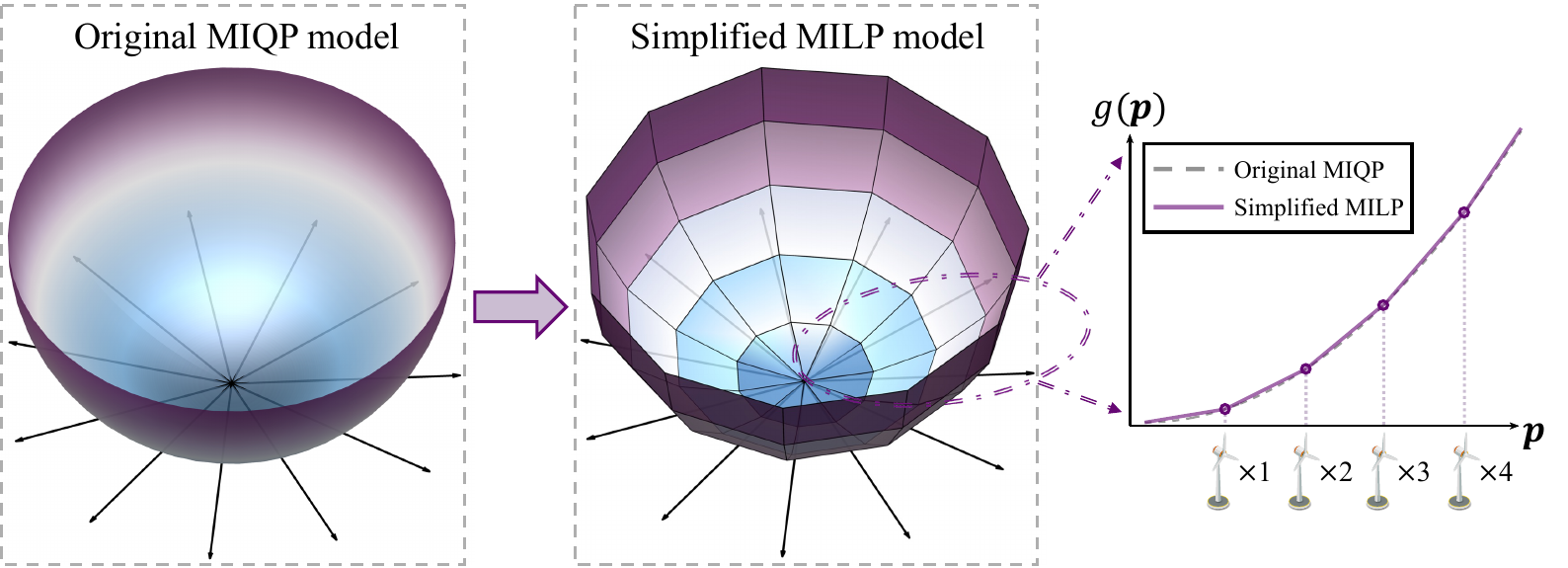}
    \caption{Illustration of A-PWL method for a nonlinear model.}
    \label{fig_Linear}
\end{figure}

It is worth noting that if the wake effects and output uncertainties of each CET are ignored and all CETs are assumed to operate at rated output, the proposed linearization becomes exact and introduces no approximation error.
The method can also be extended to multi-dimensional spaces. Taking the three-dimensional case as an example, the corresponding linearization is illustrated in Fig.~\ref{fig_Linear}. Different colors indicate distinct linear segments within the domain $[p_k, p_{k+1}]$. The black arrows represent the spatial coordinates (longitude and latitude) of different submarine cables along with their associated active power flows, while the hemispherical surface depicts the linearized values of the squared power flow.
\vspace{-1.0em}
\vspace{-2em}
\subsection{Equivalent Formulation of Absolute Value Variables}
Due to the physical constraints of the CS, the sign of $P_{ij,e}$ represents the direction of power flow. To simplify the absolute value term $\left|P_{ij,e}\right|$ in cons. \eqref{eq_Linear5}, the continuous auxiliary variable $P^{+}_{ij,e}$ is introduced to represent $\left|P_{ij,e}\right|$, and the following constraints are imposed to ensure its validity:
\begin{flalign}
P^{+}_{ij,e} \geq P_{ij,e},
&\label{eq_Linear7}
\end{flalign}
\vspace{-2em}
\begin{flalign}
P^{+}_{ij,e} \geq -P_{ij,e}.
&\label{eq_Linear8}
\end{flalign}

Since the proposed model is formulated as a minimization problem, the objective function inherently drives the power loss term $P_{ij,e}^2$ to be as small as possible, i.e., it minimizes the linear combination of $w_i$. As the auxiliary variable $P^{+}_{ij,e}$ is also a linear combination of $w_i$, the optimization naturally ensures that $P^{+}_{ij,e}$ attains the value $P^{+}_{ij,e} = \left|P_{ij,e}\right|$ in both optimal and feasible solutions.

Therefore, the compact form of the CEF-CS model is,
\begin{flalign}
\min _{z,p} \,\, \eqref{eq_Obj},
\notag
\end{flalign}
\vspace{-2em}
\begin{flalign}
\text{s.t.} \,\, \eqref{eq_topology1}-\eqref{eq_Ope3},\,\eqref{eq_Linear1}-\eqref{eq_Linear8}.
\notag
\end{flalign}

\vspace{-0.5em}
The original MIQP problem can be reformulated as an MILP problem, which is more amenable to efficient solution by modern branch-and-cut solvers.
\vspace{-1em}
\section{Case Studies}
\begin{figure}[htbp]
    \centering
    \includegraphics[width=0.9\linewidth]{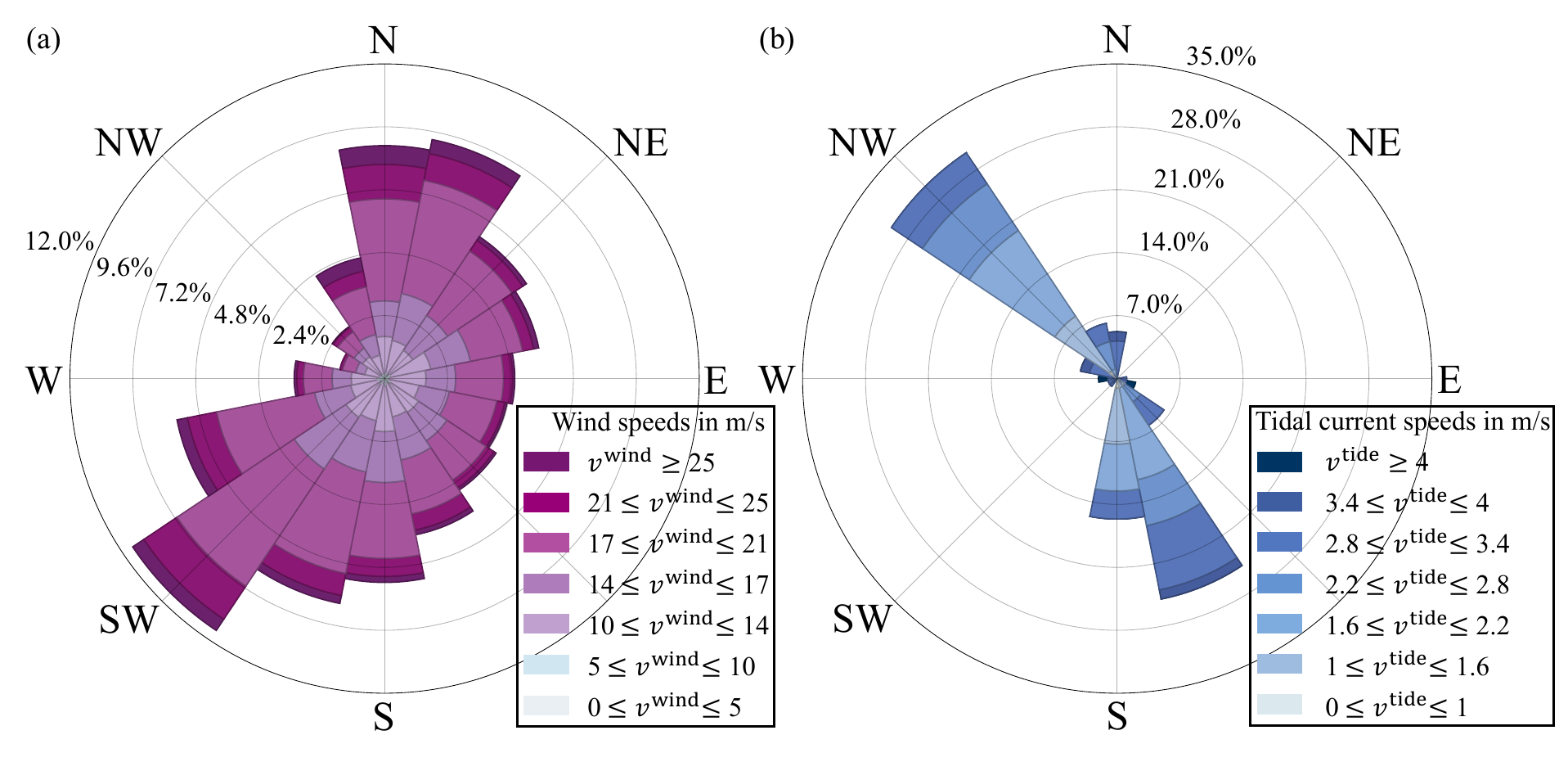}
    \caption{Wind and tidal current rose diagram.}
    \label{fig_Rose}
\end{figure}

The proposed model is tested on three CEF-CSs comprising 22, 42, and 125 CET units, respectively. The CS voltage level is set to 66 kV. The annual profiles of wind and tidal current direction and speed are sourced from the Orkney Islands, UK, and illustrated using wind rose and tidal rose diagrams in Fig. \ref{fig_Rose}, where sectors denote flow directions and concentric circles indicate the probability distribution of velocity magnitudes. 

\begin{table*}[htbp]
  \centering
  \caption{Detailed Parameters of the Selected Cables}
  \vspace{-1em}
  \label{tab:cable}
  \begin{tabular}{ccccccc}
  \hline\hline
  Cable Specifications & \makecell[c]{Cost of lazy-wave /\\ fully-suspended cable \\ ($10^{4}$ CNY/km)}  & \makecell[c]{Cost of buoyancy, bending,\\ and connection devices\\($10^{4}$ CNY/km)}  & \makecell[c]{Resistance \\($25^{\circ} \mathrm{C}$, $\Omega$/km)} & \makecell[c]{Reactance \\($25^{\circ} \mathrm{C}$, $\Omega$/km)} & \makecell[c]{Ampacity\\(A)} & \makecell[c]{Failure rate\\(times/km)}  \\ \hline
  Type 1
  &185 / 203  &55.2  &0.2460  &0.1350 &303 &0.1486  \\ 
  Type 2
    &229 / 251 &64  &0.1328  &0.1220 &424 &0.1065 \\ 
  Type 3     &280 / 308  &72.8  &0.0819   &0.1130 &534 &0.0837  \\ 
  Type 4  &355 / 390 &92  &0.0491  &0.1050 &592 &0.0648 \\ 
  \hline\hline
  \end{tabular}
\end{table*}

\begin{figure}[htbp]
    \centering
    \includegraphics[width=0.75\linewidth]{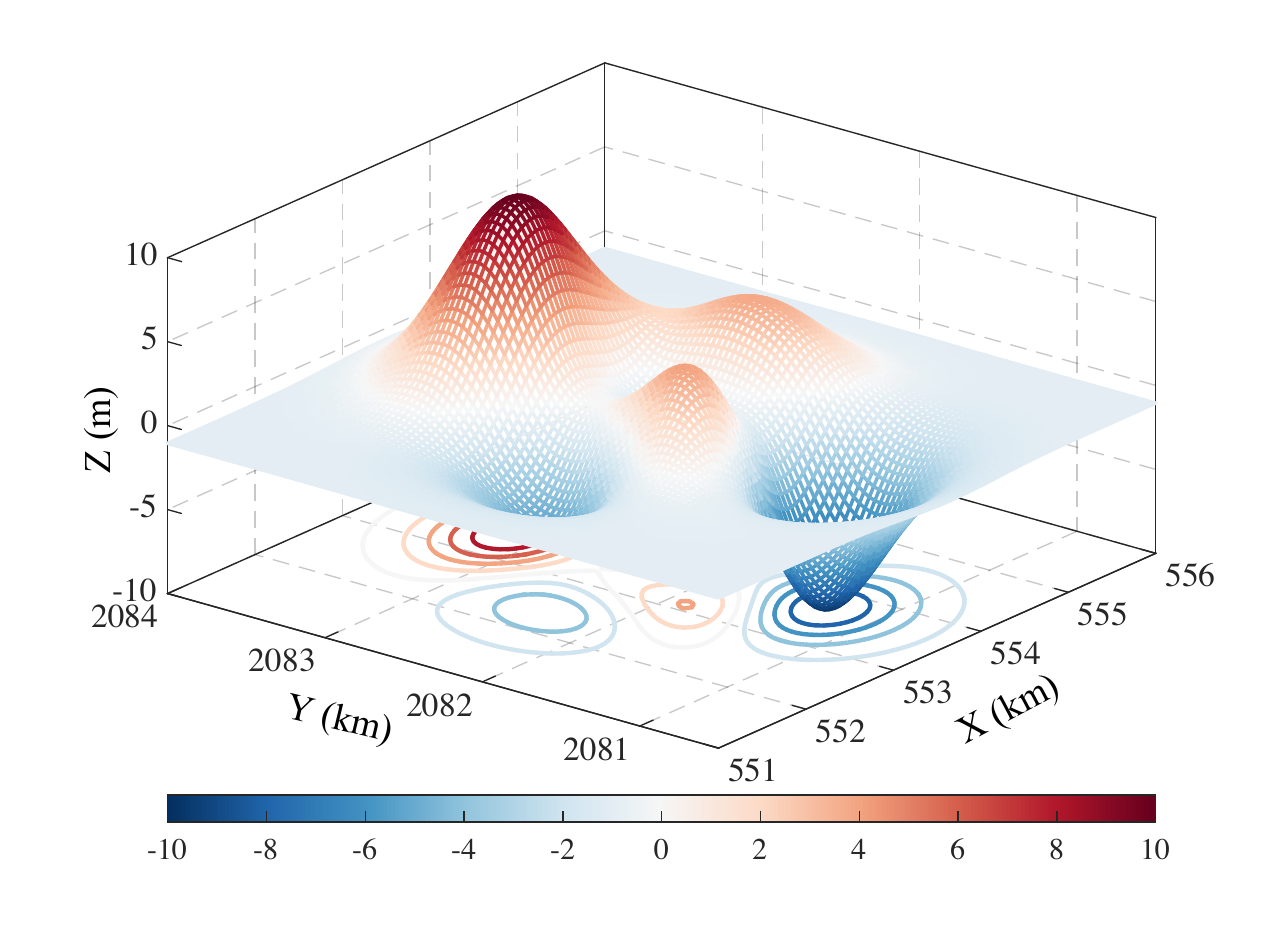}
    \caption{3D visualization of seabed topography.}
    \label{fig_Seabed}
\end{figure}

The CET parameters are specified as follows: For WT, the rotor radius is 63 m, with cut-in, rated, and cut-out wind speeds set at 3 m/s, 11.4 m/s, and 25 m/s, respectively. The rated power is 12 MW, with a thrust coefficient $C_\mathrm{T}$ of 0.8 and expansion factor $\varepsilon$ of 0.0646. For TCT, the blade radius is 10 m, the respective tidal current speeds are 0.7, 3, and 5 m/s, and rated power is 1.5 MW. The corresponding $C_\mathrm{T}$ and $\varepsilon$ are 0.7 and 0.04. The rated capacity of PV system installed at each node is set to 1 MW. Synthetic seabed topography is generated based on the geographic coordinates of CETs. The average seabed depth is 120 m, with elevation bounds $\underline{h}$ and $\overline{h}$ set to ±5 m, as illustrated in Fig. \ref{fig_Seabed}. The CEF is assumed to operate 3000 hours annually over a 20-year design life, with an inflation rate of 0.08. The feed-in tariff for renewable energy is 0.5 CNY/kWh, while for conventional thermal units it is 0.45 CNY/kWh. The regional carbon price is 50 CNY/t, with emission intensity of 850 g/kWh. CS operation assumes the mean time to repair of 1200 hours, with the repair rate coefficient of 912.5, the maintenance coefficient $k_\mathrm{m}$ of 0.01, and installation cost coefficient $r_\mathrm{c}$ of 1.5 \cite{lerch2021metaheuristic}. The vessel rental cost, auxiliary transport cost and mobilization cost are 2 million CNY/day, 1.3 million CNY,  and 3 million CNY. All submarine cables use copper conductors, with a density of 8.9 t/m³ and recycling price of 36{,}696 CNY/t. The recycling transport coefficient $k_\mathrm{d}$ is 0.15. Cable design margins are defined as 1.05 for static lazy-wave sections, 1.391 for dynamic segments, and 1.5 for fully-suspended configurations. Detailed specifications are listed in Table \ref{tab:cable}. By default, each OSS is designed to accommodate a maximum of 8 outgoing feeders, with a maximum receivable active power of 330 MW if the OSS-capacity constraint is activated. The CAPEX for the tower, foundation, and generator is fixed at 14.727 million CNY/MW. For OWF and TCF, the CS accounts for 18\% of the CAPEX. In CEF, while tower and foundation costs remain consistent (26\% of CAPEX), the CS accounts for 21\%\cite{lande2019co}.

The model was formulated using the YALMIP tool in MATLAB (2024a) and evaluated with the GUROBI Optimizer (12.0.0) on the Apple M3 Pro (12-core CPU, 18-core GPU).

\vspace{-1em}
\vspace{-1em}

\vspace{-1.2em}
\subsection{Simulation Results}

To validate the effectiveness of the proposed CET model for multi-energy complementary generation from various deep-sea renewable energy sources, four case studies are outlined:

\textit{Case A1}: CS planning for 22 proposed CETs

\textit{Case A2}: CS planning for 22 WTs

\textit{Case A3}: CS planning for 22 TCTs

\textit{Case A4}: CS planning for 22 co-located WTs and TCTs

\begin{figure*}[htbp]
    \centering
    \includegraphics[width=0.86\textwidth]{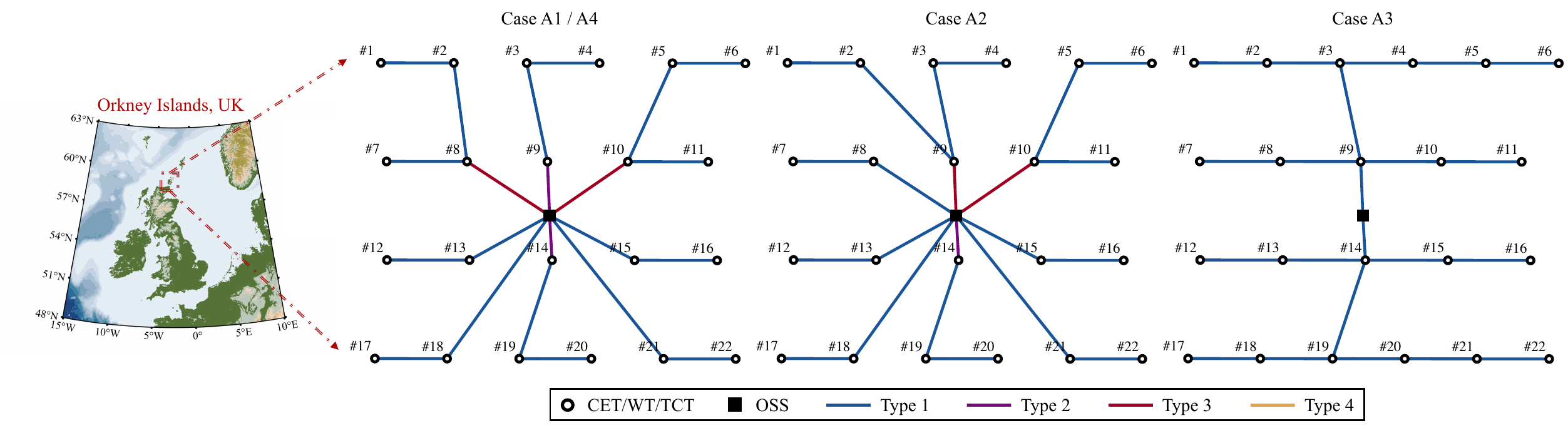}
    \caption{Planning results for cases of different renewable energy sources.}\label{fig_A}
\end{figure*}

The CS planning results and the corresponding cost breakdowns for the multi-renewable generation farm across the four cases are presented in Fig. \ref{fig_A} and Table \ref{tab:comparison results A}.

\begin{table}[htbp]
  \centering
  \caption{Annualized Economic comparison of Different Offshore Renewable Energy Sources($10^{4}$ CNY \textyen)}
  \label{tab:comparison results A}
  \begin{adjustbox}{max width=\columnwidth}
  \begin{tabular}{ccccc}
  \hline\hline
  Description & \textit{Case A1*} & \textit{Case A2} & \textit{Case A3} & \textit{Case A4} \\ \hline
  WT                        &$\surd$ &$\surd$ &---  &$\surd$      \\ 
  TCT                       &$\surd$ &---  &$\surd$     &$\surd$  \\
  PV                       &$\surd$ &---  &---     &---   \\\hline
  Investment cost           & 1011.46 & 1004.91 & 767.78 & 1011.46  \\ 
  Construction cost           & 436.63 & 437.45 & 406.24 & 436.63  \\ 
  Maintenance cost           & 10.11 & 10.05 & 7.68 & 10.11  \\ 
  Decommissioning cost           & 50.27 & 51.05 & 52.73 & 50.27  \\ 
  Operation cost         & 285.05 & 221.75  & 11.47  & 242.29     \\ 
  Reliability cost     & 4.53 & 3.75 & 0.37 & 3.98     \\ 
  Proactive curtailment cost  & 0.00 & 0.00 & 0.00 & 0.00     \\ 
  Total cost of CS         & 1798.05 & 1728.97 & 1246.28  & 1754.74    \\ \hline
  LCOE (\textyen/MWh)        & 157.41 & 220.60 & 2083.08  & 178.29    \\
  \hline\hline
  \end{tabular}
  \end{adjustbox}
\end{table}
In terms of the planning results, Case A1 and Case A4 yield identical topology results. Except for Case A3, the remaining three cases adopt eight feeders and utilize the first three cable types (Type 1–Type 3), with Type 1 cables being predominantly selected. Type 2 and Type 3 cables are only used for connections involving the OSS. In Case A3, as only TCTs are deployed with relatively low rated power per turbine, all submarine cables are of Type 1 to minimize the total cost of CS, and the number of feeders is reduced to four.

In terms of CS cost comparison, Case A1 incurs the highest total cost, while Case A3 has the lowest. Given that Case A1 and Case A4 share the same CS topology, their investment, construction, maintenance, and decommissioning costs are identical. However, due to the inclusion of PV generation at each turbine in Case A1, its operation cost and reliability cost increase by 17.6\% and 13.8\%. Case A2 and Case A3 consider only WT and TCT generation, respectively, the results indicate that the higher the total active power output per node, the greater the overall CS cost. However, when extending the analysis from CS cost to the overall LCOE of the renewable energy system, Case A1 demonstrates the highest economic efficiency, followed by Case A4. This highlights the economic benefits of multi-energy complementarity in deep-sea renewable energy systems: the integration of more renewable energy types leads to improved cost-effectiveness. Conversely, Case A3 shows that a standalone TCF system offers limited cost-performance advantages. This is primarily because the compact spatial layout of turbines leads to pronounced wake effects, which significantly reduce the actual output of tidal current energy. Overall, the WT–TCT–PV complementary generation scheme emerges as the most economically viable and promising configuration.

To further compare and analyze the impact of different dynamic cable configurations and OSS constraints on the proposed multi-scenario CEF-CS planning, eight additional cases are designed as follows:

\textit{Case B1}: CS planning for 22 CETs with lazy-wave cables under multi-scenario uncertainty

\textit{Case B2}: CS planning for 22 CETs with fully-suspended cables under multi-scenario uncertainty

\textit{Case B3}: CS planning for 22 CETs with lazy-wave cables without considering multi-scenario uncertainty

\textit{Case B4}: CS planning for 22 CETs with lazy-wave cables under multi-scenario uncertainty and tightened OSS feeders restriction

\textit{Case B5}: CS planning for 22 CETs with lazy-wave cables under multi-scenario uncertainty and the tightest OSS feeders restriction

\textit{Case B6}: CS planning for 22 CETs with lazy-wave cables under multi-scenario uncertainty and maximum active power limit of OSS

\textit{Case B7}: CS planning for 22 CETs with lazy-wave cables under multi-scenario uncertainty and intra-day PV fluctuation

\textit{Case B8}: CS planning for 22 CETs with lazy-wave cables under multi-scenario uncertainty using a Taylor-series-based loss approximation adapted from \cite{park2021optimal}

\begin{figure}[htbp]
    \centering
    \includegraphics[width=0.9\linewidth]{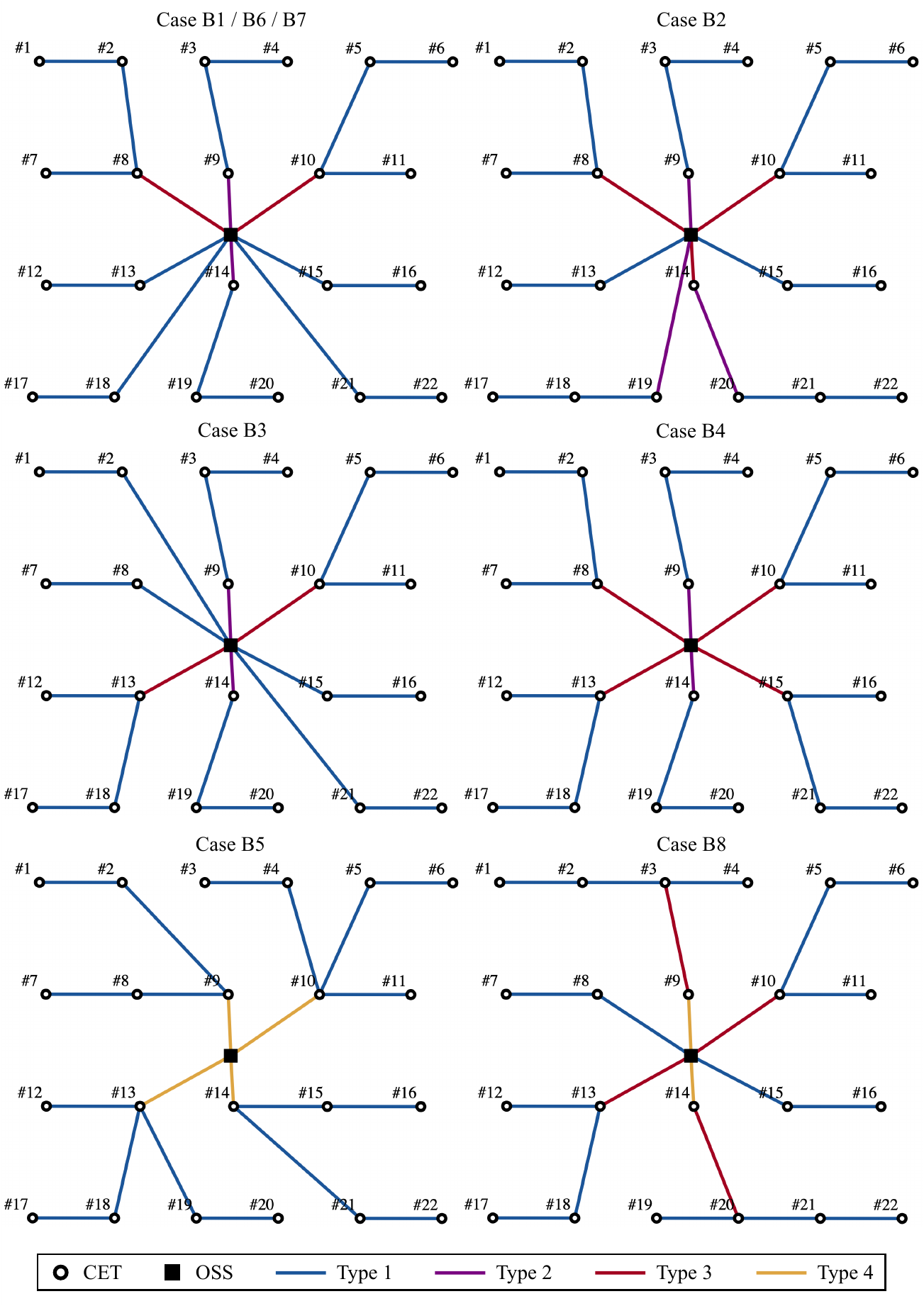}
    \caption{Planning results for cases of different cable configurations.}
    \label{fig_B}
\end{figure}

Cases B1 and B2 both incorporate multi-scenario renewable energy outputs and adopt lazy-wave and fully-suspended cable configurations for CS planning, respectively (The settings of Case B1 are identical to those of Case A1).  Case B3 assumes all renewable sources operate at their rated capacities, thereby ignoring wake losses and output fluctuations across different scenarios. Case B4 modifies B1 by tightening the constraint on OSS, reducing the maximum number of allowable feeders to six, to simulate more stringent practical requirements for OSS in CEF-CS planning. 
Case B5 further reduces the maximum number of OSS feeders to four, so that the physical and economic impact of a severe feeder restriction can be observed more clearly. Case B6 imposes a maximum receivable active power limit on the OSS to test whether proactive curtailment becomes active under a transfer bottleneck. Case B7 introduces intra-day PV fluctuation to examine the influence of PV output uncertainty. Case B8 replaces the proposed A-PWL loss approximation with a Taylor-series-based approximation to provide an alternative linearization benchmark adapted from \cite{park2021optimal}.
The planning results and associated cost breakdowns are shown in Fig. \ref{fig_B} and Table \ref{tab:comparison results B}.

In terms of the planning results, most cases mainly adopt the first three cable types, while the stricter feeder-limited case further activates the highest-capacity Type-4 cable near the OSS. Cases B1, B6, and B7 share the same topology, indicating that the OSS active-power limit and PV fluctuation mainly affect operating costs rather than the radial layout in this test system. Case B2 uses seven feeders, with a Type-2 submarine cable between \#14 and \#20 CETs to reduce total cost. Cases B3 and B8 show local changes in OSS-connected feeders and cable sizing because the deterministic rated-output assumption and the Taylor-series loss approximation alter the evaluated cable-loading pattern. Across the cases, the number of CETs per feeder is evenly distributed, typically ranging from two to five units. In Cases B4 and B5, the allowable count of OSS feeders is lowered from eight to six and four individually. Consequently, more CETs have to be clustered together using inter-CET connecting cables. For the 22-CET radial CS, the number of inter-CET cables increases from 14 in Case B1 to 16 in Case B4 and 18 in Case B5, which explains the stronger branching structure and the use of higher-capacity root branches in Fig. \ref{fig_B}.

\begin{table*}[htbp]
  \centering
  \caption{Annualized Economic comparison of Different Cable Configurations ($10^{4}$ CNY \textyen)}
  \label{tab:comparison results B}
  \begin{tabular}{ccccccccc}
  \hline\hline
  Description & \textit{Case B1*} & \textit{Case B2*} & \textit{Case B3} & \textit{Case B4} & \textit{Case B5} & \textit{Case B6} & \textit{Case B7} & \textit{Case B8}\\ \hline
  Lazy-wave  &$\surd$ &--- &$\surd$  &$\surd$  &$\surd$ &$\surd$ &$\surd$ &$\surd$\\ 
  Fully-suspended  &--- &$\surd$  &---   &---   &--- &--- &--- &---\\
  Multi-scenario &$\surd$ &$\surd$ &---  &$\surd$   &$\surd$ &$\surd$ &$\surd$ &$\surd$\\ 
  PV intra-day fluctuation &--- &--- &---  &---   &--- &--- &$\surd$ &---\\ 
  OSS maximum power limit &--- &--- &---  &---   &--- &$\surd$ &--- &---\\ 
  OSS feeders limit ($\leq$)  & 8 & 8 & 8 & 6 & 4 & 8 & 8 & 8  \\
  Approximation method  & A-PWL & A-PWL & A-PWL & A-PWL & A-PWL & A-PWL & A-PWL & Taylor  \\
  \hline
  Investment cost           & 1011.46 & 1013.52 & 1013.15 & 1002.96  & 1094.58 &1011.46 &1011.46 &1011.31 \\ 
  Construction cost           & 436.63 & 164.06 & 437.20 & 429.05  & 427.96 &436.63 &436.63 &420.80 \\ 
  Maintenance cost           & 10.11 & 10.14 & 10.13 & 10.03  & 10.95 &10.11 &10.11 &10.11 \\ 
  Decommissioning cost           & 50.27 & 3.30 & 50.47 & 46.80  & 43.28 &50.27 &50.27 &42.38 \\ 
  Operation cost         & 285.05 & 308.64  & 609.58  & 331.06  & 428.11 &269.17 &253.88 &44.98   \\ 
  Reliability cost     & 4.53 & 1.26 & 8.24 & 4.19  & 4.03 &4.53 &4.15 &3.80    \\ 
  Proactive curtailment cost  & 0.00 & 0.00 & 0.00 & 0.00  & 0.00 &73.87 &0.00 &0.00    \\ 
  Total cost of CS         & 1798.05 & 1500.92 & 2128.77  & 1824.08  & 2008.90 &1856.05 &1766.50 &1533.38   \\ 
  \hline\hline
  \end{tabular}
\end{table*}

The comparison of CS cost components can be seen in Table \ref{tab:comparison results B}. When considering both wake effects and multi-scenario output variability of the CEF, Case B2 with fully-suspended cables incurs slightly higher investment and operation costs than Case B1 with lazy-wave cables, by \(2.06\times10^4\) CNY and \(2.36\times10^5\) CNY, respectively. However, due to the absence of buoyancy and connector devices required for lazy-wave configurations, Case B2 achieves much lower construction and decommissioning costs, ultimately reducing the total CS cost by 16.5\%. In Case B3, all renewable generators are assumed to operate at rated power, disregarding wake losses and output fluctuations. This changes the CS topology and significantly increases cable loading, leading to a \(3.25\times10^6\) CNY rise in operation cost and an 18.4\% increase in total cost compared with Case B1. Cases B4 and B5 jointly reveal the impact of OSS feeder restrictions. In Case B4, reducing the maximum feeder number from eight to six only increases the total CS cost by 1.4\%, because the reduction in civil cost partly offsets the increase in loss cost. However, when the feeder limit is further tightened to four in Case B5, more CETs have to be aggregated through inter-CET cables, increasing the operation cost by \(1.43\times10^6\) CNY and the total CS cost by 11.7\% compared with Case B1. Case B6 shows that proactive curtailment becomes active when the OSS power-transfer capability is constrained: although the operation cost decreases by \(1.59\times10^5\) CNY because part of the power injection is curtailed, a curtailment cost of \(7.39\times10^5\) CNY is incurred and the total CS cost increases by 3.2\%. This indicates that proactive curtailment is not triggered in the base cases because the selected cables and OSS capacity are sufficient, but it becomes effective under practical bottleneck conditions; otherwise, a passive no-curtailment strategy with a prohibitive penalty would force the planner to adopt larger and more expensive cables to collect all renewable power. Case B7 further considers intra-day PV fluctuation. The operation cost decreases by \(3.12\times10^5\) CNY and the total CS cost decreases by 1.8\% relative to Case B1, while the topology remains unchanged. This suggests that the base-case scenario construction is not dominated by the simplified PV treatment, although PV variability affects the probability-weighted cable losses. Case B8 compares the proposed A-PWL method with a Taylor-series-based loss approximation. More importantly than the large operation cost deviation itself, the Taylor approximation changes the CS topology and cable sizing relative to the A-PWL solution, while underestimating the operation cost by about 84.2\% compared with Case B1. This confirms that a local Taylor approximation is unreliable for accelerating CS planning with wide cable-loading ranges, because it cannot provide a globally valid approximation of the convex quadratic loss term. This highlights the necessity of balancing all life-cycle cost components during CS planning. Therefore, the planning of the CEF-CS must comprehensively account for wake effects, stochastic multi-source output variability, feeder restrictions, power-transfer bottlenecks, and approximation accuracy, rather than simply assuming constant rated outputs or using a non-conservative loss approximation. Moreover, if external physical risks such as anchor damage are disregarded, fully-suspended submarine cables demonstrate greater economic viability than lazy-wave configurations for deep-sea CEF applications under the adopted cost assumptions.

\vspace{-1em}
\subsection{Sensitivity Analysis of Lazy-Wave Vessel and Auxiliary Costs}

\begin{figure}[htbp]
    \centering
    \includegraphics[width=0.8\linewidth]{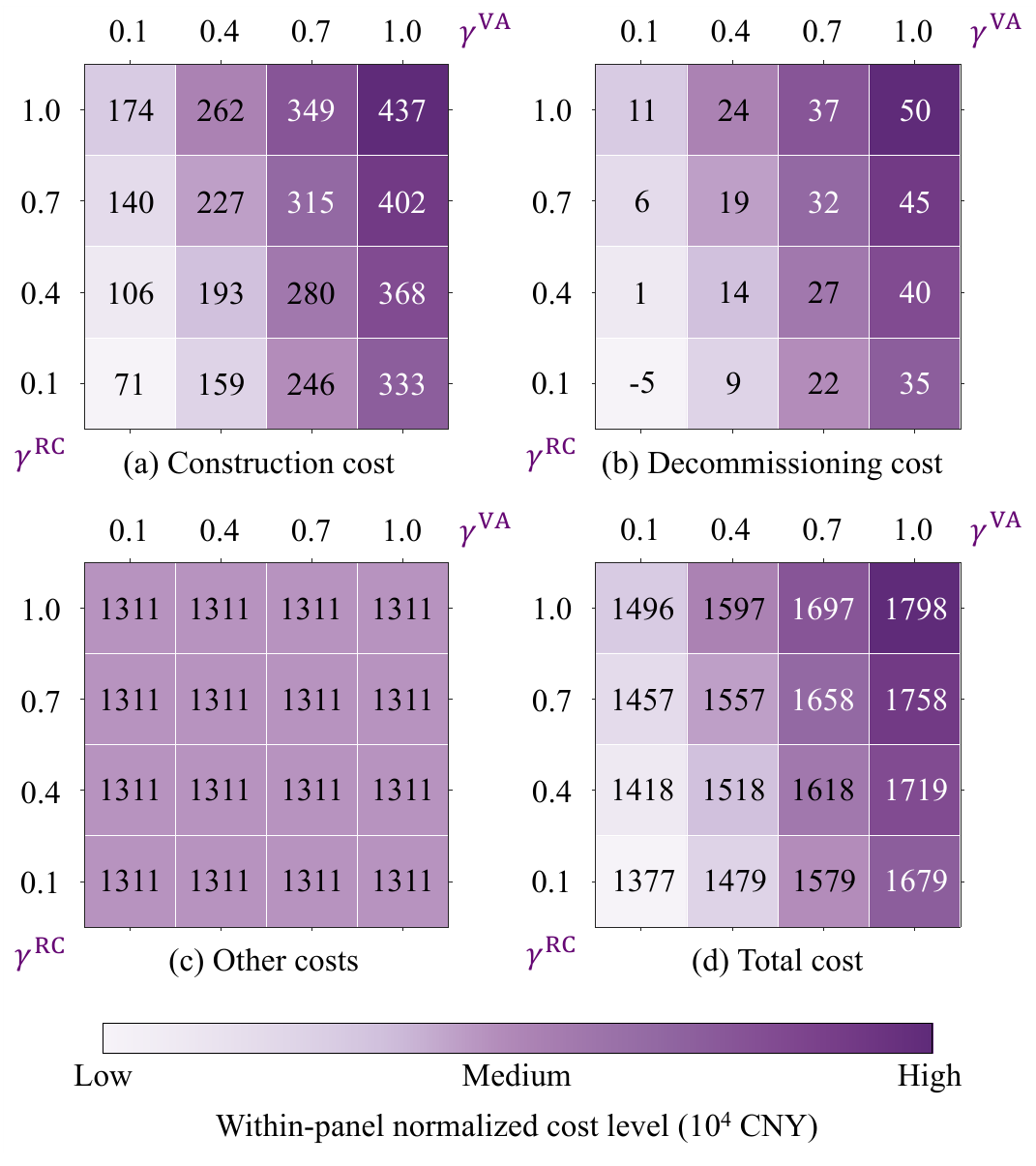}
    \caption{Sensitivity analysis of lazy-wave vessel and auxiliary costs.}
    \label{fig_lazy_wave_sensitivity}
\end{figure}

The cost advantage of fully-suspended cables over lazy-wave cables depends on the assumed construction-related coefficients of the lazy-wave configuration. To examine this dependence, a two-parameter sensitivity analysis is conducted for Case B1. Let \(\gamma^\mathrm{VA}\) denote the multiplier applied to vessel and auxiliary device related costs, and let \(\gamma^\mathrm{RC}\) denote the multiplier applied to the cable installation coefficient \(r_\mathrm{c}\). Both multipliers are varied within \(\{0.1,0.4,0.7,1.0\}\). The resulting construction cost, decommissioning cost, other costs, and total CS cost are shown in Fig. \ref{fig_lazy_wave_sensitivity}.

Fig. \ref{fig_lazy_wave_sensitivity} shows that the construction and decommissioning costs increase monotonically with \(\gamma^\mathrm{VA}\) and \(\gamma^\mathrm{RC}\), whereas the other cost components remain nearly unchanged, indicating that the topology and cable type allocation are stable and that the total-cost variation is mainly driven by vessel, auxiliary device, and installation cost assumptions. Under the baseline setting \((\gamma^\mathrm{VA},\gamma^\mathrm{RC})=(1.0,1.0)\), the lazy-wave total CS cost is \(1.798\times10^7\) CNY, about 19.8\% higher than the fully-suspended cost in Case B2. When both coefficients are reduced to the lower bound, the lazy-wave cost decreases to \(1.377\times10^7\) CNY, becoming 8.3\% lower than the fully-suspended case. Therefore, the economic advantage of fully-suspended cables should be regarded as conditional on the adopted cost assumptions, rather than as a universal engineering rule.

\vspace{-1em}
\subsection{Scenario-Number Sensitivity Analysis}

To clarify the meaning of the multi-scenario representation and to examine whether the default scenario number affects the CS planning result, a scenario-number sensitivity analysis is conducted for the benchmark multi-scenario case, i.e., Case B1. As defined in Section~II-A, a scenario \(e\in E\) denotes a representative wind--tidal joint operating state with an associated planning weight \(p_e\). Different scenario numbers (\(S=6,12,18,24,30\)) are tested using the same technical and economic parameters. The 24-scenario case is the default setting used in the main case studies, while the 30-scenario case is used as the highest-resolution reference among the tested cases. This test examines the robustness of the CS planning results to the number of representative scenarios under the adopted planning-oriented wind--tidal coupling.

  \begin{table}[htbp]
  \centering
  \caption{Annualized Economic Comparison under Different Scenario Numbers ($10^{4}$ CNY \textyen)}
  \label{tab:scenario_number_sensitivity}
  
  \begin{adjustbox}{max width=\columnwidth}
  \begin{tabular}{cccccc}
  \hline\hline
  Description & \textit{$S=6$} & \textit{$S=12$} & \textit{$S=18$} & \textit{$S=24$} & \textit{$S=30$} \\ \hline
  Investment cost              & 1011.46 & 1011.46 & 1011.46 & 1011.46 & 1011.46 \\
  Construction cost            & 436.63  & 436.63  & 436.63  & 436.63  & 436.63  \\
  Maintenance cost             & 10.11   & 10.11   & 10.11   & 10.11   & 10.11   \\
  Decommissioning cost         & 50.27   & 50.27   & 50.27   & 50.27   & 50.27   \\
  Operation cost               & 334.22  & 303.21  & 290.39  & 285.05  & 278.78  \\
  Reliability cost             & 5.07    & 4.65    & 4.57    & 4.53    & 4.46    \\
  Proactive curtailment cost   & 0.00    & 0.00    & 0.00    & 0.00    & 0.00    \\
  Total cost of CS             & 1847.77 & 1816.33 & 1803.43 & 1798.06 & 1791.72 \\  \hline
  Topology/cable type                & Same  & Same  & Same  & Default & Same  \\
  \hline\hline
  \end{tabular}
  \end{adjustbox}
\end{table}

Table~\ref{tab:scenario_number_sensitivity} summarizes the sensitivity results. The selected CS topology and cable type allocation remain identical for all tested scenario numbers, which directly confirms that the obtained CS planning result is stable with respect to the scenario resolution. Because the same cables and cable types are selected for all tested scenario numbers, the investment, construction, maintenance, and decommissioning costs remain unchanged. The slight variation in the total CS cost is mainly caused by the probability-weighted operation and reliability costs, which are evaluated under different cable-loading states. As the scenario number increases, the annual wind--tidal variability is represented more finely, and the expected operation and reliability costs gradually decrease in this tested case. The total CS cost of the default 24-scenario case is already very close to that of the 30-scenario reference, while the same CS topology and cable type allocation are retained. From the computational perspective, increasing the scenario number enlarges the scenario-indexed operation constraints and variables. Accordingly, the number of model nonzeros increases from \(1.2\times10^5\) to \(5.7\times10^5\). The branch-and-cut solution time is not strictly monotonic because it depends on presolve, incumbent discovery, cutting planes, and branching behavior. Considering both planning stability and model scale, the default \(S=24\) is adopted in the main case studies.

\begin{figure}[htbp]
    \centering
    \includegraphics[width=0.98\linewidth]{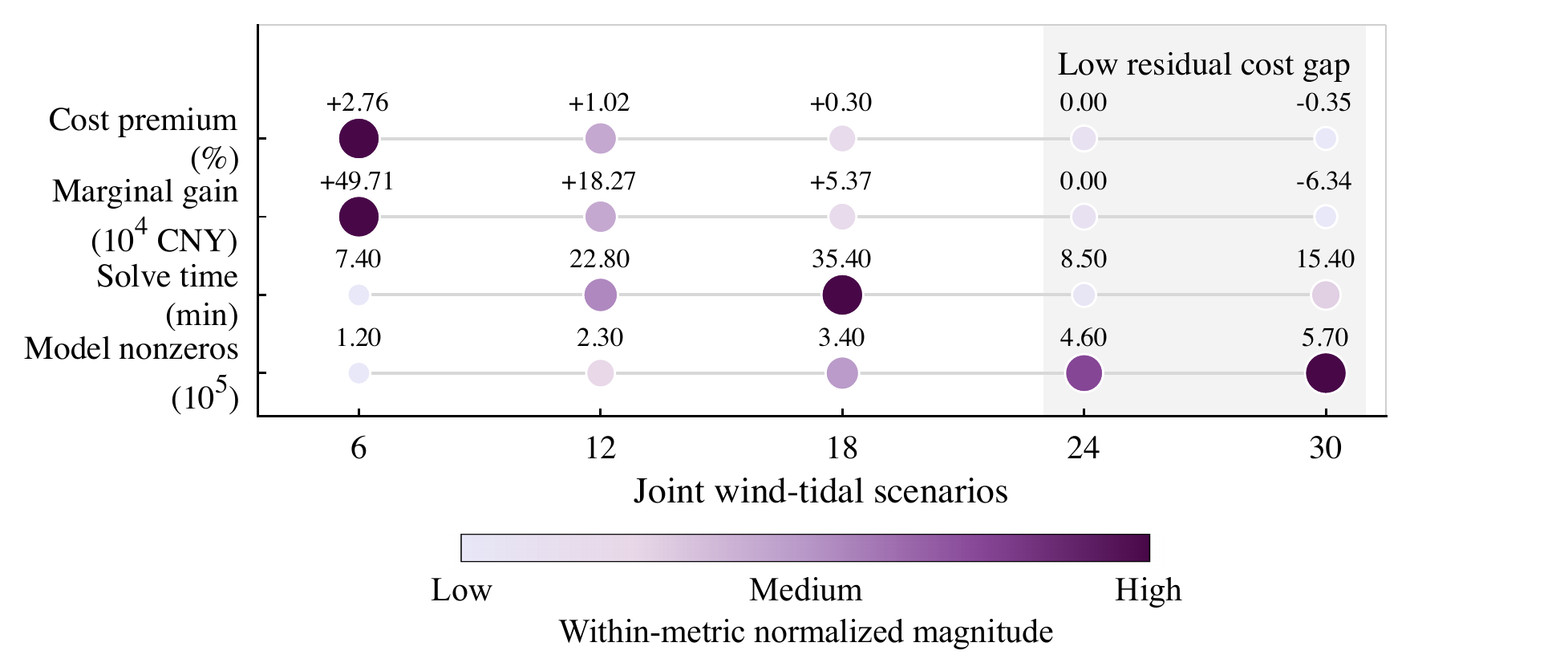}
    \caption{Sensitivity analysis of the number of joint wind--tidal scenarios.}
    \label{fig_scenario}
\end{figure}
\vspace{-0.7em}

Fig.~\ref{fig_scenario} further visualizes the tradeoff between residual cost gap and computational burden. The cost-premium row and marginal-gain row are calculated with respect to the default \(S=24\) case, and the figure acts as a streamlined decision aid.

\vspace{-1em}
\subsection{Comparative Analysis of Large-Scale Case Studies}

To further validate the effectiveness of the proposed model and methodology, the larger-scale CS case studies involving 42 CETs from \cite{Gao2026OffshoreWindCollectionMILP} and 125 CETs from \cite{yang2026optimal} are conducted. Two configurations, namely lazy-wave and fully-suspended submarine cables, are respectively applied for comparative planning.

\textit{Case C1}: CS planning for 42 CETs with lazy-wave cables under multi-scenario uncertainty

\textit{Case C2}: CS planning for 42 CETs with fully-suspended cables under multi-scenario uncertainty

\textit{Case C3}: CS planning for 125 CETs with lazy-wave cables under multi-scenario uncertainty

\textit{Case C4}: CS planning for 125 CETs with fully-suspended cables under multi-scenario uncertainty

\begin{figure}[htbp]
    \centering
    \includegraphics[width=0.98\linewidth]{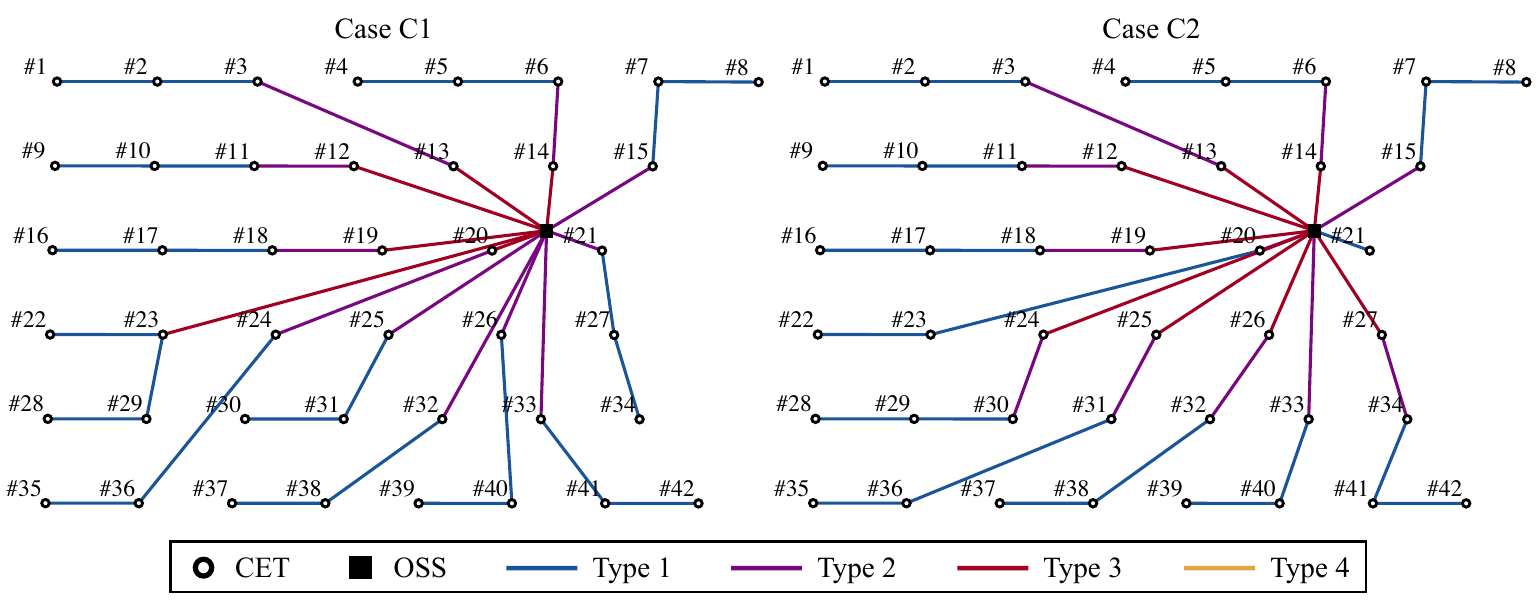}
    \caption{Planning results for a large-scale CEF with 42 CETs.}
    \label{fig_C}
\end{figure}

\begin{figure}[htbp]
    \centering
    \includegraphics[width=0.99\linewidth]{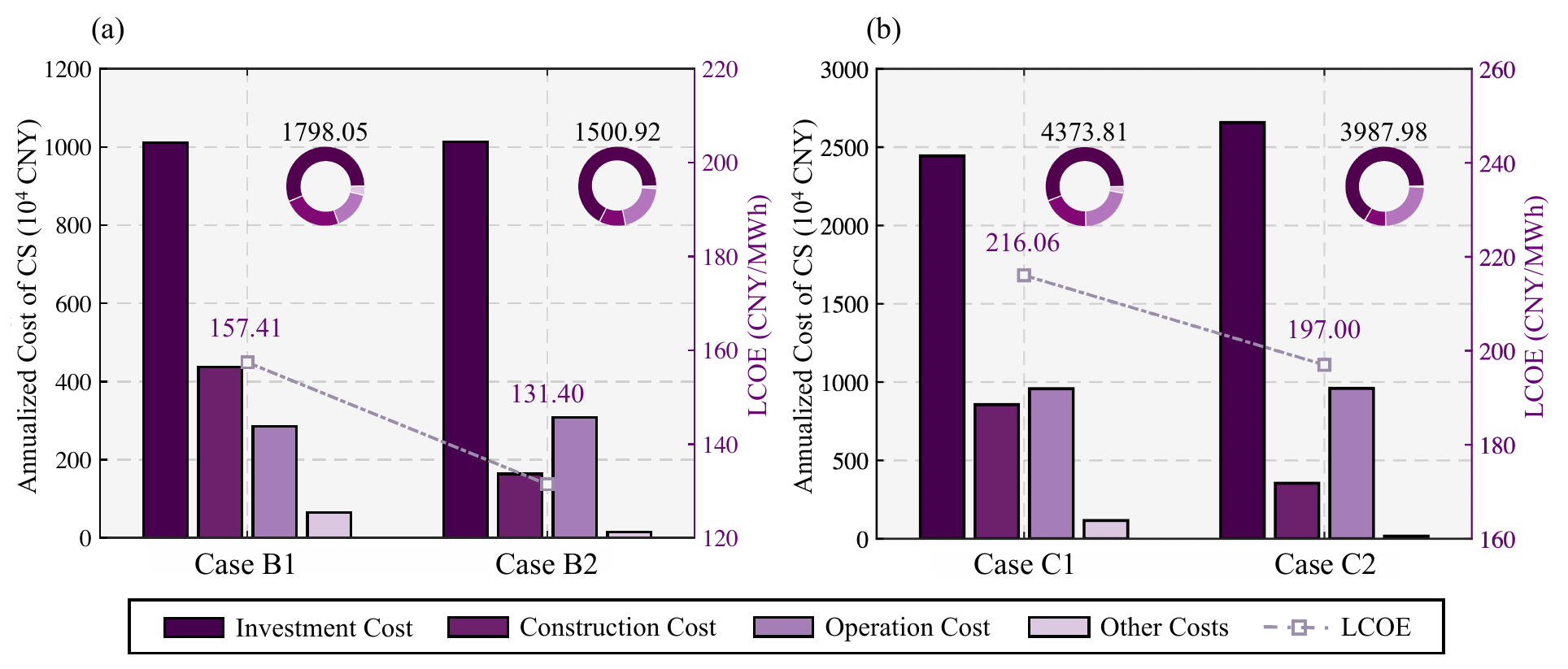}
    \caption{Cost comparison of CS across different-scale cases.}
    \label{fig_Cost}
\end{figure}

\begin{figure}[htbp]
    \centering
    \includegraphics[width=0.97\linewidth]{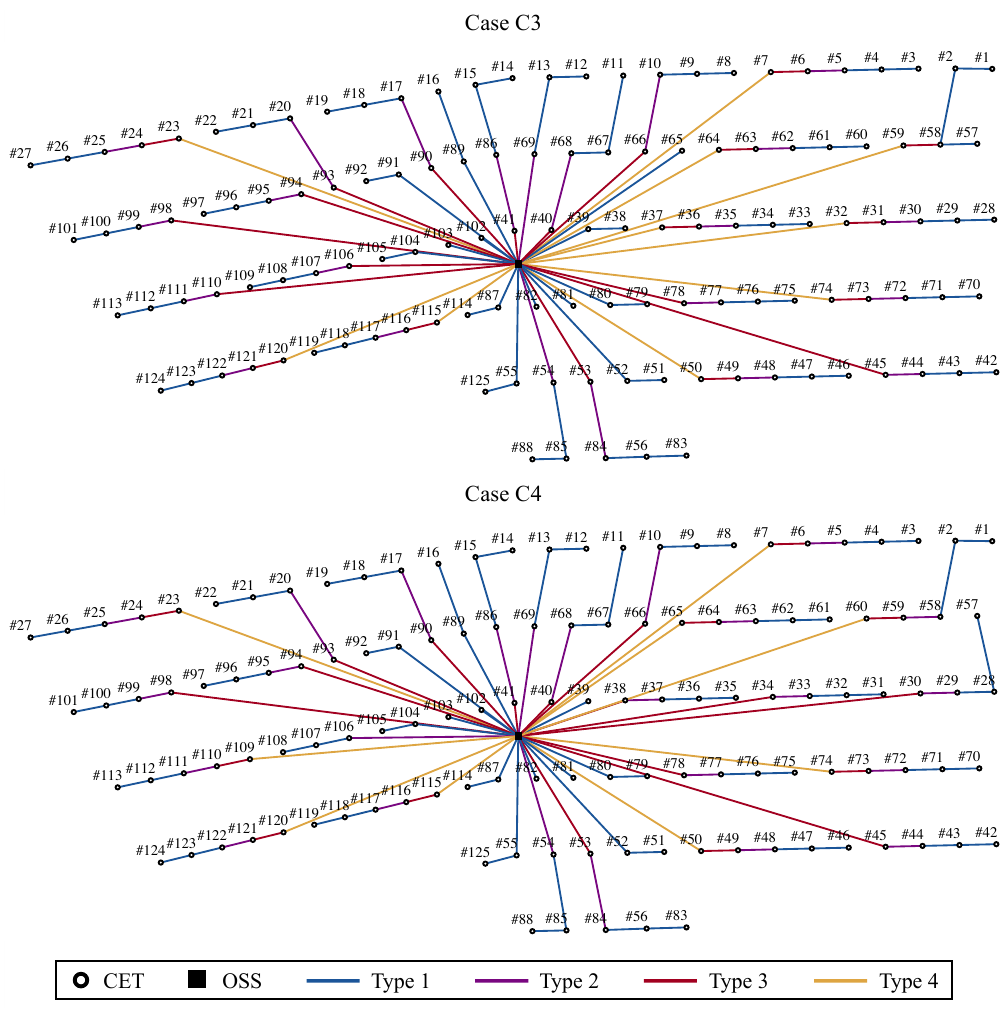}
    \caption{Planning results for a large-scale CEF with 125 CETs.}
    \label{fig_C125}
\end{figure}

The planning results and corresponding cost comparisons for the CEF with 42 CETs are presented in Fig. \ref{fig_C} and Fig. \ref{fig_Cost}. The 125-CET topologies in Fig. \ref{fig_C125} further demonstrate the scalability of the proposed planning framework. As the number of CETs increases, longer radial feeder chains and more internal aggregation branches appear. High-capacity cables are mainly selected near the OSS, where multiple downstream CET outputs are aggregated, while lower-capacity cables are mostly used at feeder tails. This allocation is consistent with the physical power-collection pattern of a radial CS. 

\vspace{-1em}
\vspace{-1.0em}

From the planning results, it is evident that as the number of CETs increases, all associated costs rise accordingly. However, for both CEF scales, the fully-suspended configuration consistently demonstrates superior economic performance over the lazy-wave configuration, whether assessed by the life-cycle cost of the CS or by the LCOE of the whole CEF. Specifically, the fully-suspended configuration reduces the total CS cost by 16.5\% in the 22-CET case and by 8.8\% in the 42-CET case. This cost advantage is mainly associated with reduced construction and decommissioning costs, and should be interpreted together with the sensitivity analysis in Fig. \ref{fig_lazy_wave_sensitivity}.

\vspace{-1.0em}

\subsection{Comparison Before and After A-PWL Method}

\begin{figure}[htbp]
    \centering
    \includegraphics[width=0.82\linewidth]{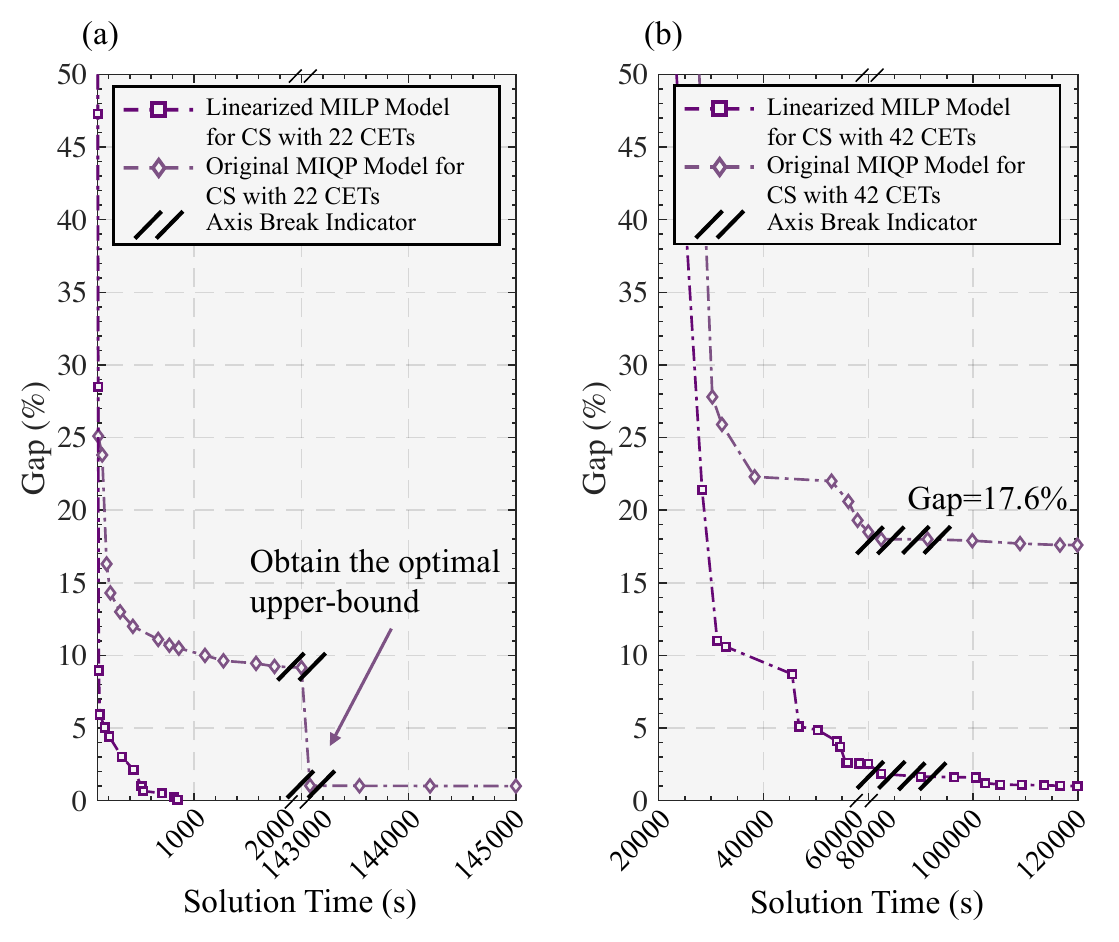}
    \caption{Solution gap of the CS planning model: (a) 22 CETs in Case B1 and (b) 42 CETs in Case C1.}
    \label{fig_Gap}
\end{figure}

To evaluate the solving accuracy and effectiveness of simplifying the large-scale MIQP problem into an MILP model using piecewise linearization, this study compares the solution time and optimality gap before and after linearization for Case B1 (22 CETs) and Case C1 (42 CETs). The comparison results are presented in Fig. \ref{fig_Gap}(a) and Fig. \ref{fig_Gap}(b), respectively.

Due to the long solving time, we consider a solution global optimal if the optimality gap falls below 1.00\%. In Case B1, the linearized MILP model confirmed optimality (gap = 0.00\%) in only 841 seconds, whereas the original MIQP model took 28,413 seconds merely to identify the optimal upper bound. It then required over 100,000 additional seconds to verify optimality. From a computational efficiency perspective, the MILP model’s solving time accounts for just 0.58\% of that required by the MIQP model. In terms of planning outcomes, the original MIQP model yields the exact same CS topology and cable type allocation as Case B1 in Fig. \ref{fig_B}. Therefore, the cost components determined by the discrete planning decisions remain unchanged, and the 1.12\% increase in total CS cost is mainly caused by the conservative overestimation of the quadratic operating loss term introduced by the A-PWL. This deviation remains within an acceptable engineering tolerance and does not affect the engineering feasibility of the CS planning result. By contrast, the Taylor-series-based benchmark in Case B8 changes both the CS topology and cable type allocation while significantly underestimating the operation cost, indicating that it does not provide a globally valid upper bound approximation of the original loss cost minimization problem. This additional comparison further supports the use of the proposed A-PWL formulation for conservative and reliable operation cost evaluation.

In Case C1, the expanded scale of CETs significantly increases the number of candidate submarine cables, leading to a power-law growth in the total number of CS planning schemes. Consequently, the solving time rises sharply compared to Case B1. The linearized MILP model achieved optimality in 119,147 seconds. In contrast, within a 120,000-second time limit, the original MIQP model failed to find the optimal upper bound, with the gap only reduced to 17.6\%. Both the planning topology and the associated costs obtained from the original MIQP model were inferior to those of the MILP model. Case studies with different CET scales show that the proposed A-PWL method significantly improves CS planning efficiency with negligible loss of physical accuracy.

\vspace{-0.1em}
\vspace{-1em}
\section{Conclusion}

This paper proposes a life-cycle planning framework for the CS of deep-sea CEFs, integrating multi-spatial wind, tidal current, and PV generation. A comprehensive economic analysis is conducted for different types of co-located renewable generation, taking into account wake effects and stochastic variability. The proposed model incorporates various CS configurations, including lazy-wave and fully-suspended dynamic submarine cables, and evaluates the investment, construction, maintenance, decommissioning, operational, and reliability costs over the entire life-cycle of the CS.

To enhance computational efficiency, the original large-scale MIQP problem is reformulated as the simplified MILP problem via the A-PWL method according to the outputs of CETs. The model further simplifies the absolute value representation of line power flows by leveraging CS-specific physical insights.

Case studies demonstrate that multi-spatial energy complementarity significantly improves the economic performance of deep-sea renewable farms. Accounting for wake effects and stochasticity is shown to influence the planning results, while comprehensive life-cycle cost modeling enables the identification of balanced and cost-effective CS planning. Under the adopted cost assumptions and without explicitly modeling external physical risks acting on dynamic submarine cables (e.g., anchor damage, fishing-gear interaction, or floating-ice impact), the fully-suspended cable configuration outperforms the lazy-wave design in terms of overall economic viability. The added sensitivity analysis shows that this conclusion is cost-condition dependent: lazy-wave cables can become competitive when vessel, auxiliary device, and installation cost coefficients are sufficiently low. The proposed linearization method maintains near-identical solution accuracy while substantially reducing computation time, thereby validating its effectiveness and feasibility. 

In conclusion, the proposed CS planning model and simplification method offer practical value and broad prospects for deep-sea CEFs. By jointly considering WT, PV and TCT, the framework provides a planning-oriented representation of multi-layer marine-space utilization, aligning with recent policy priorities for the high-quality development of the marine economy. Future work may extend this framework to include additional marine energy sources, explicitly incorporate external physical risks affecting dynamic submarine cables into the planning model, and explore advanced planning strategies, including OSS and CET micro-siting.

\bibliographystyle{IEEEtran}
\vspace{-1em}
\small\bibliography{reference}

\begin{IEEEbiography}[{\includegraphics[width=1in,height=1.25in,clip,keepaspectratio]{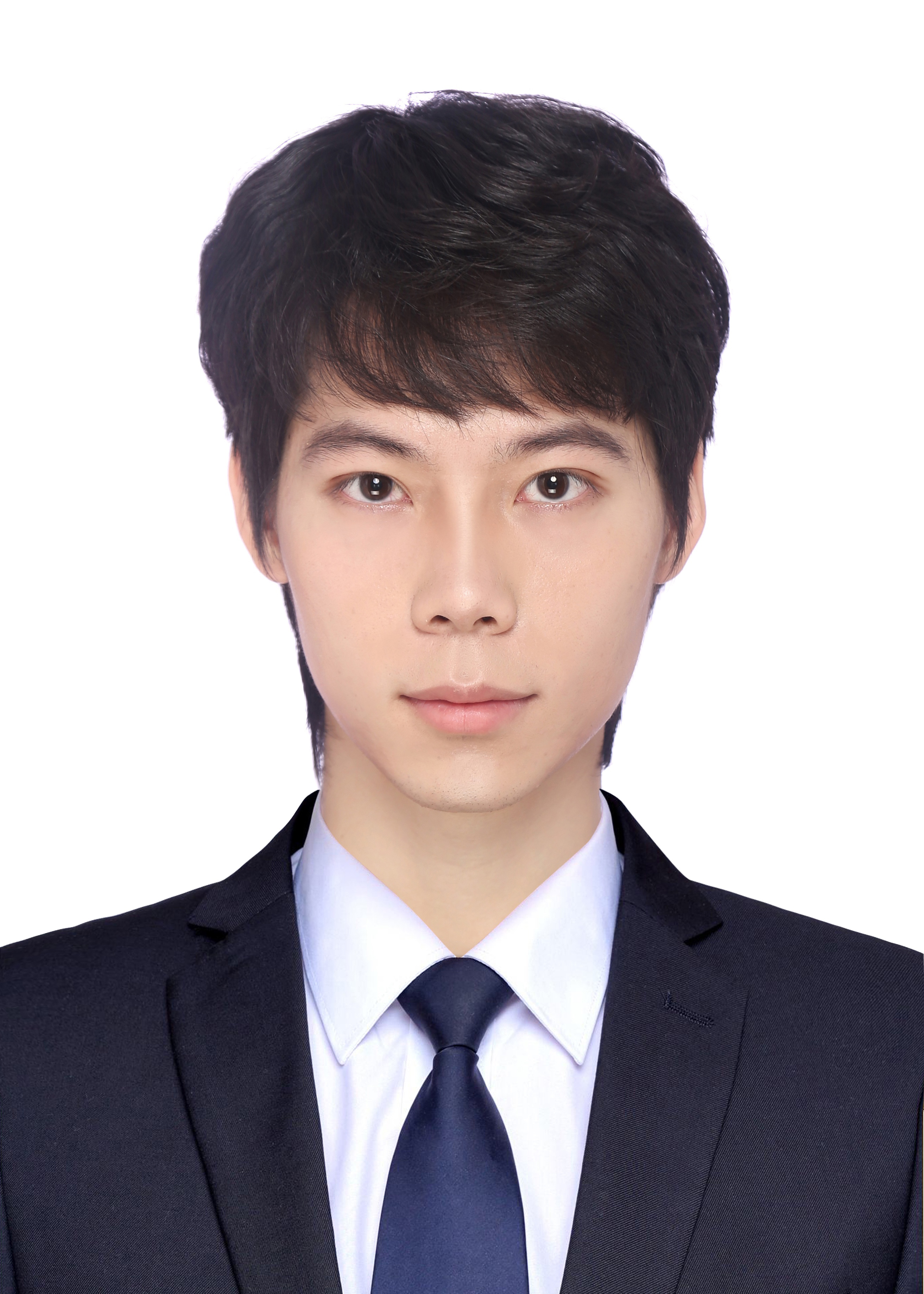}}]{Wenhao Gao}
(Student Member, IEEE) received the B.Eng. degree in electrical engineering from Chongqing University, Chongqing, China, in 2022, and the M.S. degree in electrical engineering from National University of Singapore, Singapore, in 2023. He is currently pursuing the Ph.D. degree in electrical engineering with Tsinghua University, Beijing, China. His research interests include power system planning and optimization, with emphasis on electrical collector systems for offshore renewable energy resources, coastal distribution networks, and electric-vehicle charging infrastructure.
\end{IEEEbiography}

\begin{IEEEbiography}[{\includegraphics[width=1in,height=1.25in,clip,keepaspectratio]{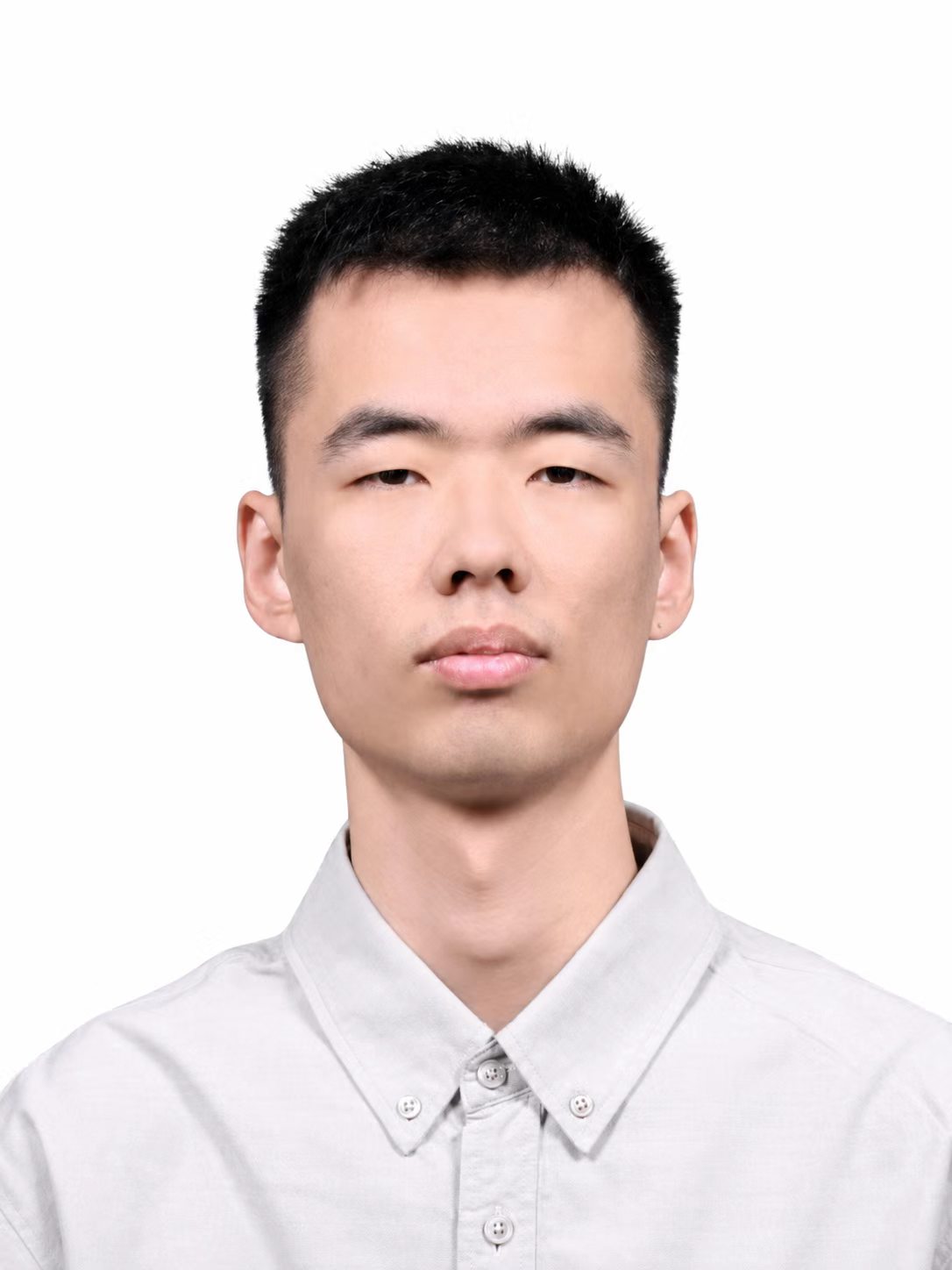}}]{Weitai Xu}
(Student Member, IEEE) received the B.Eng. degree in electrical engineering from Tsinghua University, Beijing, China, in 2025. He is currently pursuing the Ph.D. degree in electrical engineering with Tsinghua University, Beijing, China. His research interests include AI-enabled joint planning of offshore wind farm micro-siting and electrical collector systems.
\end{IEEEbiography}

\begin{IEEEbiography}[{\includegraphics[width=1in,height=1.25in,clip,keepaspectratio]{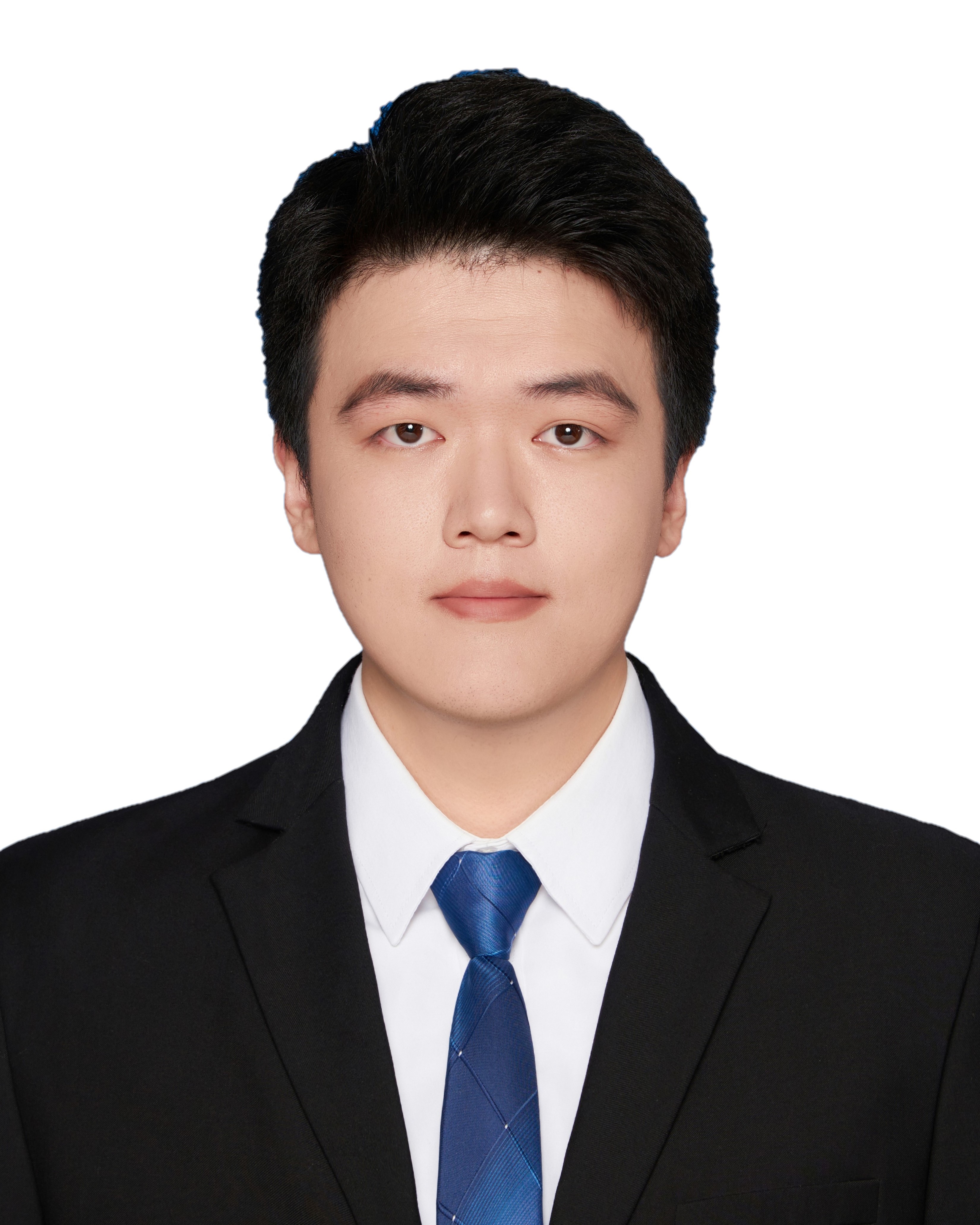}}]{Yunfei Du}
(Student Member, IEEE) received the B.Eng. degree in electrical engineering from Shandong University, Jinan, China, in 2019, and the M.S. degree in electrical engineering from the Huazhong University of Science and Technology, Wuhan, China, in 2022. He is currently pursuing the Ph.D. degree in electrical engineering with Tsinghua University, Beijing, China. His research interests include hybrid offshore wind farm optimization (integrating offshore wind farms with hydrogen or other Power-to-X systems).
\end{IEEEbiography}

\begin{IEEEbiography}[{\includegraphics[width=1in,height=1.25in,clip,keepaspectratio]{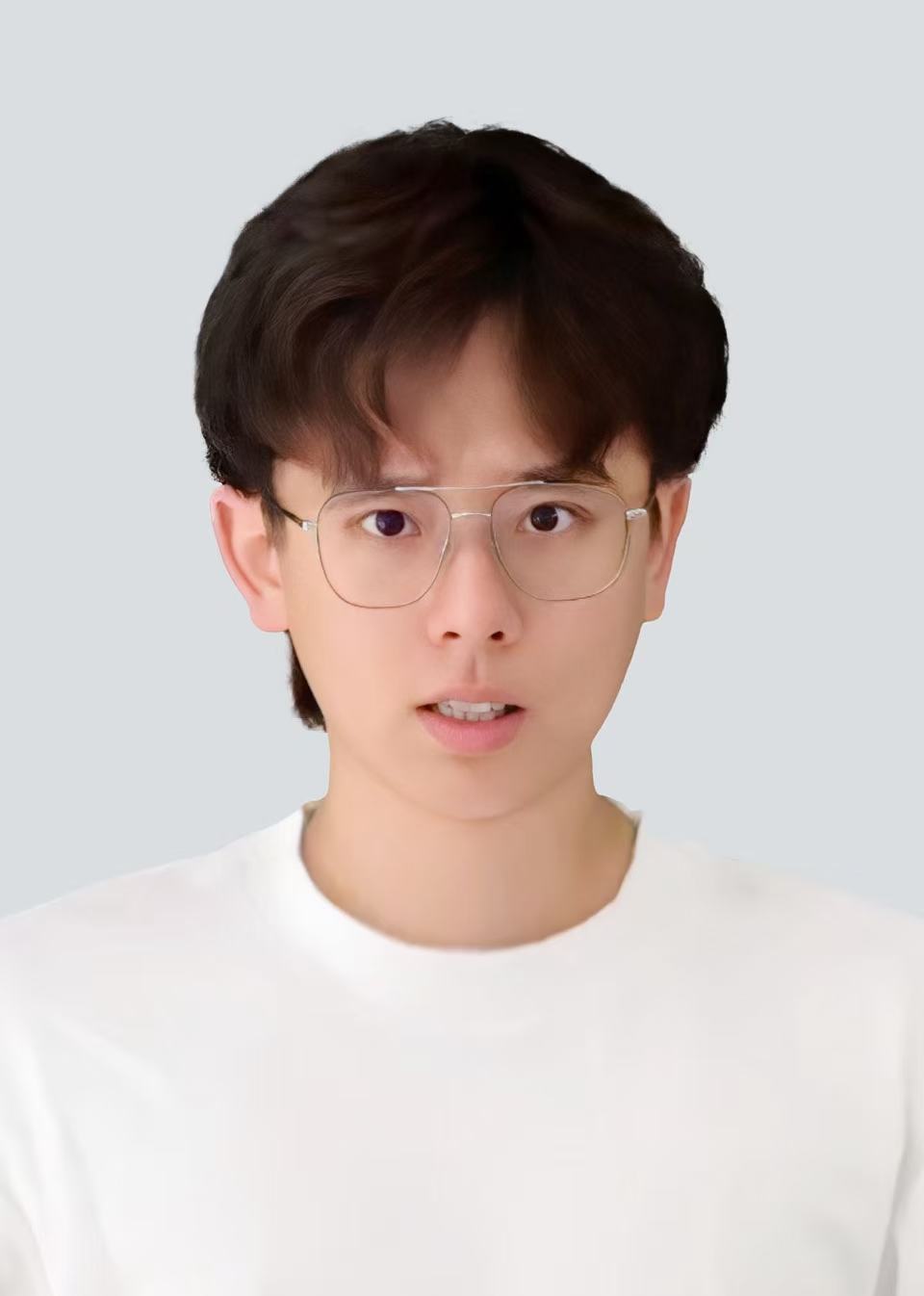}}]{Yongheng Wang}
(Student Member, IEEE) received the B.Eng. degree in electrical engineering from South China University of Technology, Guangzhou, China, in 2021, and the M.Phil. degree in electrical engineering from Tsinghua University, Beijing, China, in 2024. He is currently pursuing the Ph.D. degree in electrical engineering with The University of Hong Kong, Hong Kong SAR, China. His research interests include dissipativity-based decentralized stability criteria for converter-dominated power systems and coordinated planning of active distribution networks with electric-vehicle charging infrastructure. He was a recipient of the Best Paper Award at the 2026 IEEE PES International Meeting.
\end{IEEEbiography}

\begin{IEEEbiography}[{\includegraphics[width=1in,height=1.25in,clip,keepaspectratio]{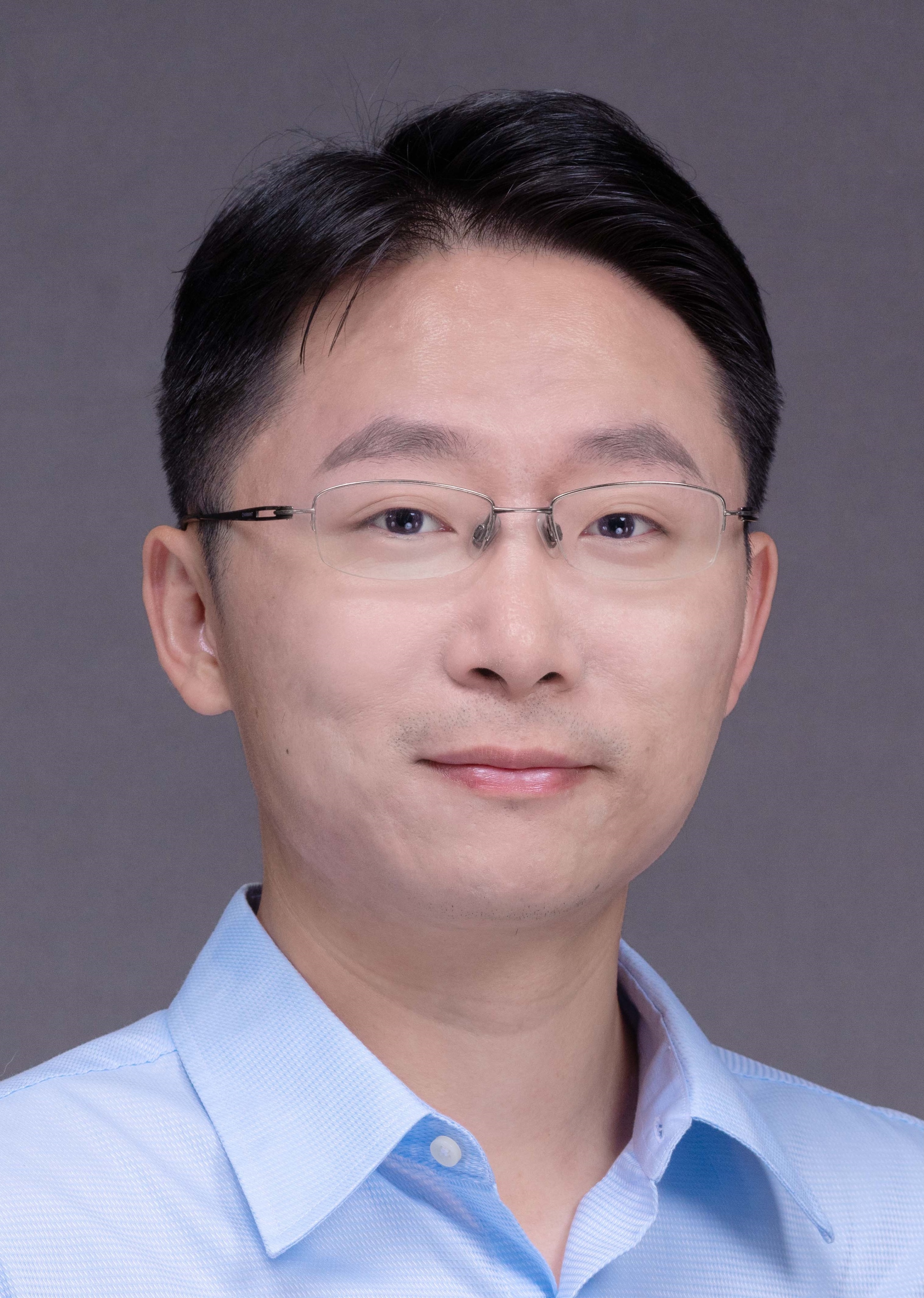}}]{Xinwei Shen}
(Senior Member, IEEE) received the B.Eng. and Ph.D. degrees from the Department of Electrical Engineering, Tsinghua University, Beijing, China, in 2010 and 2016, respectively. He was a Visiting Scholar with the Illinois Institute of Technology in 2014, the University of California at Berkeley, in 2017, and the University of Macau in 2021. He is currently an Associate Professor with Tsinghua Shenzhen International Graduate School, Tsinghua University. His research interests include Energy Internet / power systems / integrated energy systems / offshore renewable energy optimization. He is the recipient of the IEEE PES Technical Council Young Professionals Award (in 2023, the first recipient in Asia--Pacific region) and the “Young Elite Scientists Sponsorship Program” by CSEE, the Best Paper Award at IEEE PES GM,  and wins the Excellent Youth Basic Research Fund of Shenzhen (in Math). He serves as a Young Editorial Board Member of the \textit{CSEE Journal of Power and Energy Systems} / \textit{Applied Energy}, an Associate Editor of \textit{IEEE Transactions on Sustainable Energy}, and the Secretary of the IEEE PES Energy Internet Coordinating Committee.
\end{IEEEbiography}

\end{document}